\documentclass[twocolumn,preprintnumbers,amsmath,amssymb,graphicx]{revtex4}
\usepackage{graphicx}
\begin{document}
\title{Anisotropic Spherically Symmetric Collapsing Star From Higher Order Derivative Gravity Theory}
\author{HOSSEIN GHAFFARNEJAD}
\affiliation{Faculty of Physics, Semnan University, Semnan, IRAN,
 Zip code: 35131-19111}

\altaffiliation{Email address:\\hghafarnejad@yahoo.com\\
hghafarnejad@profs.semnan.ac.ir }

 \maketitle
  \textbf{\textsl{ABSTRACT}-}
  Adding linear combinations $R^2,R_{\mu\nu}R^{\mu\nu}$ and
$R_{\mu\nu\eta\delta}R^{\mu\nu\eta\delta}$ with Eins-tein-Hilbert
action we obtain interior metric of an anisotropic spherically
symmetric collapsing (ASSC) stellar cloud. We assume stress tensor
of the higher order geometrical terms to be treat as anisotropic
imperfect fluid with time dependent density function $\rho(t)$ and
radial and tangential pressures $p_r(t)$ and $p_t(t)$
respectively. We solved linearized metric equation via
perturbation method and obtained 12 different kinds of metric
solutions. Calculated Ricci and Kretschmann scalars of our metric
solutions are non-singular at beginning of the collapse for 2
kinds of them only. Event and apparent horizons are formed at
finite times for two kinds of singular metric solutions while 3
metric solutions exhibit with event horizon only with no formed
apparent horizon. There are obtained 3 other kinds of the metric
solutions which exhibit with apparent horizon with no formed event
horizon. Furthermore 3 kinds of our metric solutions do not
exhibit with horizons. Barotropic index of all 12 kinds of metric
solutions are calculated also. They satisfy different regimes such
as domain walls (6 kinds), cosmic string (2 kinds), dark matter (2
kinds), anti-matter (namely negative energy density) (1 kind) and
stiff matter (1 kind). Time dependent radial null geodesics
expansion parameter $\Theta(t)$ is also calculated for all 12
kinds of our metric solutions. In summary 4 kinds of our solutions
take absolutely positive value $\Theta>0$ which means the collapse
ended to a naked singularity but 8 kinds of our metric solutions
ended to a covered singularity at end of the collapse where
$\Theta\leq0$ and so trapped surfaces are appeared.\vskip 0.50 cm
 \textbf{\textsl{Keywords}}-Higher order derivative gravity, Anisotropic
 fluid, Spherically symmetric collapse, Non-singular models, Naked
 singularity, Domain walls, Cosmic string, Dark
matter, Anti-matter, Stiff matter, Trapped surface

\section{\label{I}Introduction}

Renormalization theory of expectation value of stress energy
tensor operator of propagating quantum fields in curved
space-times, leads to some geometrical state independent and also
state dependent corrections on right hand side (RHS) of Einstein`s
gravity equation. State independent geometrical objects are
obtained in terms of Lovelock polynomials [1,2] They are made from
higher order derivatives of the background metric $`g_{\mu\nu}`$
as $`R^2`,$ $\Box R,$ $`R_{\mu\nu}R^{\mu\nu}`$ and
$`R_{\mu\nu\gamma\delta}R^{\mu\nu\gamma\delta}`,$ where $`R`,$
$`R_{\mu\nu}`$ and $`R_{\mu\nu\gamma\delta}`$ denote as usual the
scalar curvature, Ricci and Riemann tensors respectively (see also
[3,4]).  The basic motivation for studying these \textit{Higher
Derivative gravity theories} comes from the fact that they provide
one possible approach to an as yet unknown quantum theory of
gravity [5]. However, the structure of classical solutions of
higher derivative gravity may provide a better approximation to
some metric solutions with respect to those provided by general
relativity. In four dimensions, due to the Gauss-Bonnet
topological invariant, the corresponding Lagrangian contains only
two quadratic terms among the above three possible.  Cosmological
applications of this formalism have been considered mainly in the
context of inflationary cosmology, since the R-squared term has
the virtue of inducing an early inflationary stage in the
spatially homogeneous and isotropic Friedmann-Lemaitre-Robertson
Walker (FLRW) models (see for instance [6,7]). At the quantum
level one can see [8], where the effects of curvature-squared
terms is studied on the wave function of the closed FLRW space
time by the Hawking and Luttrell. They obtained that Weyl-squared
term in the action plays no role while the $R^2$ term behaves like
a massive scalar field. The wave function of the universe can be
interpreted as corresponding in the classical limit to a family of
solutions that start out with a long period of exponential
expansion and then go over to a matter dominated era. Also
Wheeler-DeWitt wave equation of quantum cosmology with $R^2$ term
is solved by Kasper [9] (see also [10]). The quantum cosmology of
introducing a cubic term in the scalar curvature into the
Einstein-Hilbert action was studied previously in the ref. [11].
 In the present work we want to study physical effects of the Lovelock polynomials on evolution of ASSC stellar cloud as follows.
\\
 We use now the Einstein-Hilbert action added by combinations of the geometrical
objects $`R_{\mu\nu}R^{\mu\nu}`,$
$`R_{\mu\nu\gamma\delta}R^{\mu\nu\gamma\delta}`,$ and $R^2$ to
characterize structure of interior metric of ASSC star. Here we
assume that geometrical source of the modified Einstein`s equation
treats as classical stress tensor of anisotropic fluid stellar
matter with barotropic index $\gamma=\frac{p}{\rho}$ and
anisotropic index $\Delta=\frac{p_r-p_t}{\rho}$ where $p_r$ and
$p_t$ are radial and transverse pressures. $\rho$ and $p$ are
energy density and isotropic pressure.  We solved linearized
dynamical metric equation and obtained family of time dependent
metric solutions (12 different kinds) and corresponding energy
density and all pressures of the collapsed object. Mathematical
derivations show that the stellar collapse reaches to covered
singularity space time in 7 kinds of our solutions while 5 kinds
of them reach to a naked singularity which in the latter cases
cosmic censorship conjecture is violated. 10 kinds of our
solutions are asymptotically flat but 2 kinds take flat Minkowski
form at beginning of the collapse. In the latter 2 cases Ricci and
Kretschmann scalars are calculated as regular while in the former
10 ceases they become singular at beginning of the collapse.
 For 2 kinds of our metric solutions the event and apparent horizons are formed at finite
 times.4 (3) kinds of our metric solutions exhibit with event (apparent) horizon
 only at finite times. 3 kinds of our metric solutions do not
 exhibited with both event and apparent horizons.
 Details of the work are given as follows.\\
In section II, we call modified Einstein-Hilbert action functional
with additional Lovelock polynomials. In section III, we describe
summary of physical properties of an spherically symmetric
collapsing star and obtained time dependent, linearized
gravitational field equations of anisotropic spherically symmetric
time-dependent curved space-time. We assume higher order
derivative geometrical counterparts treat as anisotropic imperfect
fluid. We obtained 12 different kinds of metric solutions and
calculate time dependent energy density, radial, transverse,
isotropic and anisotropic pressures, barotropic and anisotropy
indexes of collapsing cloud. In section IV we study dynamics of
event and apparent horizon formation and also obtained equation of
trapped surfaces. In section V we study conditions on radial null
geodesics expansion parameter and obtained exactly time dependence
of them for all 12 kinds of our metric solutions. Also their
diagrams are plotted against collapsing time in figures
$1,2,3,\cdots 12$. Also our numerical results are collected at
tables 1, 2, 3, 4 and 5. Section VI denotes to concluding remarks.
\section{\label{II}Effective gravity theory}
Let us we start with the following action functional in which we
used units $`G=c=\hbar=1`$ [3,4,5]. \begin{equation}
I=\frac{1}{16\pi}\int dx^4\sqrt{g}\{R+\zeta R^2+\eta
R^{\mu\nu}R_{\mu\nu}+\xi
R^{\mu\nu\alpha\beta}R_{\mu\nu\alpha\beta}\} \end{equation}  where
the coupling constants $`\zeta`,$ $`\eta`$ and $`\xi`$ come from
dimensional regularization of interacting quantum matter fields.
They have dimensions as $(lenght)^2$ and must be determined by
experiment. Hence we solve dynamical field equations against
arbitrary values of these parameters and obtain some physical
statements. In particular case $\xi=1=\zeta$ and
 $\eta=-4$
 the above action up to term of Ricci scalar $R$ become a topological invariant (called the Euler number) and in case
 $\zeta=\frac{1}{2},\eta=-2,\xi=1$
it leads to the well known Weyl-squared scalar
$C^{\mu\nu\eta\delta}C_{\mu\nu\eta\delta}.$  $`g`$ is
absolute value of determinant of the metric field $`g_{\mu\nu}`.$\\
Varying (1), with respect to $`g^{\mu\nu}`,$ we obtain the metric
field equation as (see [3,4,5] and references therein):
\begin{equation}
 G_{\mu\nu}=8\pi T_{\mu\nu}=-(\alpha H^{(1)}_{\mu\nu}+\beta
H^{(2)}_{\mu\nu})\end{equation} where we defined
\begin{equation}\alpha=\zeta-\xi,~~~~~~\beta=\eta+4\xi,\end{equation}\begin{equation}
H_{\mu\nu}^{(1)}=2(\nabla_{\mu}\nabla_{\nu}R+RR_{\mu\nu})-g_{\mu\nu}(2\square
R+\frac{1}{2}R^2)\end{equation} and  \begin{equation}
H^{(2)}_{\mu\nu}=\nabla_{\mu}\nabla_{\nu}R-\square
R_{\mu\nu}+2R^{\alpha\beta}R_{\alpha\beta\mu\nu}$$$$-\frac{1}{2}g_{\mu\nu}(\square
R+R^{\alpha\beta}R_{\alpha\beta}). \end{equation}
 From trace of
the metric equation (2), one can obtain a good condition as
$(6\alpha+2\beta)\Box R+R+4\beta R^{\alpha\beta}R_{\alpha\beta}=0$
which help us to rewrite the metric field equation (2) as other
form. Stress tensor given in RHS of the metric equation (2) is
geometrical counterpart state independent of quantum matter field
gravitational source. It will be considered to be treat as
anisotropic imperfect fluid. We solve (2) in the next section to
obtain interior metric of ASSC star and effective energy density,
all pressure components, barotropic and anisotropy indexes of
geometrical fluid.

\section{Spherically symmetric collapsing star}

A massive star may be drop below the \textit{Chandrasekhar} or the
\textit{Oppenheimer-Volkoff} limits and so it will be collapsed
[12,4]. A proper treatment of gravitational collapse would be
prohibitively complicated because of its spherical symmetry
breaking in the presence of nonzero
 space-components of four-velocity of the perfect fluid stellar matter.
  Even for spherically symmetric configurations in which equation of state of the ASSC star $`p(\rho)`$
 is still stable as $`\frac{dp}{d\rho}\geq0`,$
  the possible end of the collapse depends on the in-falling sound velocity of the perfect
  fluid stellar matter $v_s$. For a perfect gas the equation of state is $p=\rho_m RT=\rho_mv_s^2$ in which $\rho_m$ is energy density, $R$
  is the particular gas constant, and $v_s$ is the sound velocity of the gas. It is really
  a characteristic thermal speed of the fluid molecules as
  $v_s=\sqrt{RT}.$ In case of could gas (dust), the fluid molecules move as non-relativistic and one can obtain barotropic index of state equation as
   $\gamma=\frac{p}{\rho}=\frac{\rho_mv_s^2}{\rho_m c^2}=\frac{v_s^2}{c^2}\approx0,$
where $\rho=\rho_mc^2$ and $v_s<<c,$ for a "cold" gas, and $c$ is
speed of light. In case of non-degenerate, ultra-relativistic gas
(radiation but also matter in the very early universe) the
barotropic index become $\gamma=\frac{1}{3}$ because of
$v_s\approx\frac{c}{\sqrt{3}}$ in each direction of 3-space.
Usually barotropic index $\gamma=\frac{p}{\rho}$ is called as
\textit{$\gamma$-law} condition on the state equation
$p=\gamma\rho$ which at Zel`dovich interval  $0<p<\rho$ [13] must
be satisfied as
\begin{equation} 0<\gamma(t)<1.\end{equation} In the latter situation the mean
free path between particle collisions is much less than the scales
of physical interest, then the fluid may be treat as perfect.
 This is also a good approximation to the behavior of
any form of non-relativistic fluid or gas.  In other words, dust
dominance of the stellar perfect fluid is only a hydrostatic fluid
which is rest with respect to a comoving coordinate system. In
case stiff matter of imperfect fluids the corresponding barotropic
index is $\gamma=1$ and for supper stiff matter $\gamma>1.$  In
case $\gamma=-1$ the fluid is treated as dark energy (the
cosmological constant) which can be support acceleration of cosmic
inflation. More generally, the expansion of the universe is
accelerating for any equation of state $\gamma<-\frac{1}{3}$ which
corresponds to dark energy. If $\gamma<-1$ the dark energy lies in
the phantom regime and would cause a big rip. If
$-1<\gamma<-\frac{1}{3}$ the dark energy lies in the quintessence
regime. Using the existing data, it is still impossible to
distinguish between phantom and quintessence. Critical barotropic
index $\gamma=-\frac{1}{3}$ characterizes cosmic string and causes
to expands the universe without acceleration. The latter situation
can be seen from well known Friedmann equation
$\frac{\ddot{a}}{a}=-\frac{4\pi G}{3}(1+3\gamma)\rho.$ In case
$-\frac{1}{3}<\gamma<0$ the fluid lies in the dark matter regime. Particular value $\gamma=-\frac{2}{3}$ characterizes domain wall. \\
 In order to get some feeling for what can happen
during collapse, it is considered usually the simplest case named
as \textit{Tollman} model [14,15]. In that case, the stellar
collapse is assumed to be spherically symmetric inhomogeneous
perfect fluid matter. In spherically symmetric space time, without
loss of generality, the line element can be written in diagonal
form as
\begin{equation}
ds^{2}=-e^{2\nu(t,r)}dt^2+e^{2\lambda(t,r)}dr^2+R^2(t,r)(d\theta^2+\sin^2\theta
d\varphi^2).\end{equation}  The proper radius from  center of the
collapsing fluid  is $R(t,r)$ and the collapsing boundary surface
$\Sigma$ is given in the interior comoving coordinates as a free
fall surface $r=r_0=const$ so that $R_{\Sigma}=R(t,r_0).$
 Usually the exterior metric of the
collapsing star is assumed to be static and satisfies the standard
Darmois-Israel junction conditions
(see [16,17] and reference therein).\\
We know that the geometry inside the collapsing star is dynamic
since the radial coordinate become time-like and the metric is
time dependent (see also [18]). Hence we
 choose a non-comoving observer where the interior metric of the ASSC star become
 \begin{equation} ds^2= -e^{a(t)}dt^2+e^{b(t)}dr^2+t^2e^{c(t)}(d\theta^2+\sin^2\theta d\varphi^2)\end{equation}
where $a(t), b(t), c(t)$ are determined by solving the metric
equation (2). It should be pointed that 2-sphere spatial part of
the above metric is inhomogeneous because of absence of $r^2$
term. Inserting (8) the components of the tensor equation (2)
become nonlinear and so one must be decide to solve (2) by
applying one of numerical or perturbative analytical methods. We
will consider here the latter method by assuming the following
perturbation series expansions to linearize the metric field
equation (2).
\begin{equation} e^{a(t)}= e^{a_0}\{1+\epsilon
a_1(t)+O(\epsilon^2)\},\end{equation}\begin{equation} e^{b(t)}=
e^{b_0}\{1+\epsilon b_1(t)+O(\epsilon^2)\}\end{equation} and
\begin{equation} e^{c(t)}= e^{c_0}\{1+\epsilon
c_1(t)+O(\epsilon^2)\}\end{equation} where $a_0,b_0,c_0$ are
constants and $\epsilon$ is a suitable dimensionless order
parameter of the series expansion. For instance we can choose $
\epsilon=\frac{\alpha}{\beta}$ for $\alpha<\beta$ or
$\epsilon=\frac{\beta}{\alpha}$ for $\alpha>\beta$ respectively.
In case $\alpha=\beta$ the equation (3) leads to $\zeta-\eta=5\xi$
for which we can choose
$\epsilon=\{\frac{\zeta}{\eta},\frac{\zeta}{\xi},\frac{\eta}{\xi}\},$
for $\{\zeta<\eta,\zeta<\xi,\eta<\xi\}$ respectively or vice
versa. Applying (9), (10) and (11) zero order approximation of the
metric equation (2.2) leads to the following condition.
\begin{equation}e^{a_0}=-e^{c_0}\end{equation}
where the line element (8) will be take the following form.
\begin{equation} ds^2\simeq-e^{a_0}(1+\epsilon a_1(t))dt^2+e^{b_0}(1+\epsilon b_1(t))
dr^2$$$$-t^2e^{a_0}(1+\epsilon c_1(t))(d\theta^2+\sin^2\theta
d\varphi^2).
\end{equation} First we set \begin{equation}a_0=i\pi,~~~b_0=0\end{equation} because interior metric of the collapsing
star has Euclidean signature $(+,+,\\+,+).$ Applying (9), (10),
(11), (12) and (14), first order part of nonzero
$tt,rr,\theta\theta,$ components of the metric equation (2) become
respectively
\begin{equation} (8\alpha+3\beta)\dddot{c_1}+(24\alpha+11\beta)\ddot{c_1}/t+(1-2\beta/t^2)\dot{c_1}-(1/t$$$$+(16\alpha+10\beta)/t^3)c_1
-(8\alpha+3\beta)\ddot{a_1}/t-(16\alpha+6\beta)\dot{a_1}/t^2
$$$$-(1/t-(16\alpha+10\beta)/t^3)a_1$$$$+4(2\alpha+\beta)\ddot{b}/t+(1-4(2\alpha+\beta)/t^2)\dot{b_1}=0,\end{equation}
\begin{equation}(4\alpha+\beta)\ddddot{c_1}+(20\alpha+5\beta)\dddot{c_1}/t+((4\alpha+\beta)/t^2+1)\ddot{c_1}+$$$$(2(4\alpha+\beta)/t^3
+3/t)
\dot{c_1}+(1/t^2+2(4\alpha+\beta)/t^4)c_1$$$$-(4\alpha+\beta)\dddot{a_1}/t-(4\alpha+\beta)\ddot{a_1}/t^2
+(1/t-(8\alpha+2\beta)/t^3)\dot{a_1}$$$$+(1/t^2-(8\alpha+2\beta)/t^4)a_1+(2\alpha+\beta)\ddddot{b_1}+(12\alpha+4\beta)\dddot{b_1}/t=0
\end{equation}
and
\begin{equation}(4\alpha+3\beta)\ddddot{c_1}+(40\alpha+20\beta)\dddot{c_1}/t+(2(4\alpha+13\beta)/t^2+1)\ddot{c_1}$$$$
+(2(5\beta-4\alpha)/t^3 +2/t)
\dot{c_1}-2(4\alpha+\beta)c_1/t^4+(4\alpha+\beta)\ddddot{b_1}$$$$+2(8\alpha+3\beta)\dddot{b_1}/t
+(4\beta/t^2+1)\ddot{b_1}+(2\beta/t^3+1/t)\dot{b_1}$$$$-(8\alpha+4\beta)\dddot{a_1}/t-(8\alpha+12\beta)\ddot{a_1}/t^2$$$$-
(1/t-(8\alpha-6\beta)/t^3)\dot{a_1}+(8\alpha+2\beta)a_1/t^4=0.\end{equation}
$\varphi\varphi$ component of the  metric equation (2) leads to
 the equation (17) and dose not give us more information about the metric solutions. Over dot denotes to
  differentiations with respect to time parameter
$`t`.$\\
 If we want to know about time dependence of energy density and pressures of the collapsing cloud,  we must be have Einstein tensor components
 which up to second order terms $O(\epsilon^2),$ are obtained as
 \begin{equation} G^t_t(t)\simeq-\epsilon\{(\dot{c_1}+\dot{b_1})/t+(c_1-a_1)/t^2)\},\end{equation}
 \begin{equation} G^r_r(t)\simeq-\epsilon \{\ddot{c_1}+(3\dot{c_1}-\dot{a_1})/t+(c_1-a_1)/t^2\},\end{equation}
 and \begin{equation} G^{\varphi}_{\varphi}=G^{\theta}_{\theta}\simeq-\epsilon\{(\ddot{b_1}+\ddot{c_1})/2+(2\dot{c_1}+\dot{b_1}-\dot{a_1})/2t
\}\end{equation} where we insert (12) and (15). Anisotropy
property of our gravitational system can be seen from inequality
between (19) and (20) as $G^r_r\neq
G^{\theta}_{\theta}=G^{\varphi}_{\varphi}.$ Hence we assume that
right side of the equation (2) describes an imperfect fluid
stellar matter source with anisotropic stress tensor
\begin{equation}
T_{\mu}^{\nu}=diag[-\rho(t),p_r(t), p_{t}(t), p_{t}(t)]
\end{equation}
where $p_r(t)$ and $p_t(t)$ are radial and transverse (tangential)
pressures respectively. We decompose the stress tensor (21) to two
parts as
\begin{equation} T_{\mu}^{\nu}=diag[-\rho,p+\Pi_{r}^{r}, p+\Pi_{\theta}^{\theta}, p+\Pi_{\varphi}^{\varphi}]\end{equation}
where $p(t)$ is isotropic pressure as
\begin{equation} p(t)=\frac{p_r+2p_t}{3}\end{equation}
and $\Pi_{\mu}^{\nu}$ is  traceless anisotropic stress tensor as
\begin{equation}\Pi_{\mu}^{\nu}(t)=(p_r-p_t)diag\bigg[0,\frac{2}{3},-\frac{1}{3},-\frac{1}{3}\bigg]\end{equation}
which are determined from trace of the tensors (21) and (22).
Applying (18), (19), (20) and comparing (21) with (2), one can
obtain
\begin{equation} \rho(t)\simeq\frac{\epsilon}{8\pi}\{(\dot{c_1}+\dot{b_1})/t+(c_1-a_1)/t^2)\}\end{equation}
\begin{equation} p_r(t)\simeq-\frac{\epsilon}{8\pi}\{\ddot{c_1}+(3\dot{c_1}-\dot{a_1})/t+(c_1-a_1)/t^2\},\end{equation}
\begin{equation}p_t(t)\simeq-\frac{\epsilon}{8\pi}\{(\ddot{b_1}+\ddot{c_1})/2+(2\dot{c_1}+\dot{b_1}-\dot{a_1})/2t
 \}\end{equation}
\begin{equation}p(t)\simeq-\frac{\epsilon}{24\pi}\{2\ddot{c_1}+\ddot{b_1}+5\dot{c_1}/t+\dot{b_1}/t-\dot{a_1}/t)\}.\end{equation}
We have also dimensionless anisotropy index
\begin{equation}\Delta(t)=\frac{(p_t-p_r)}{\rho}\simeq\{\ddot{c_1}/2-\ddot{b_1}/2+2\dot{c_1}/t-\dot{b_1}/2t-\dot{a_1}/2t$$$$
+(c_1-a_1)/t^2
\}/\{(\dot{c_1}+\dot{b_1})/t+(c_1-a_1)/t^2\}\end{equation} and
dimensionless barotropic index
\begin{equation} \gamma(t)=\frac{p(t)}{\rho(t)}\simeq-\frac{1}{3}\{2\ddot{c_1}+\ddot{b_1}+5\dot{c_1}/t+\dot{b_1}/t
-\dot{a_1}/t\}/\{(\dot{c_1}+\dot{b_1})/t$$$$+(c_1-a_1)/t^2\}.\end{equation}
 If
 we want to know about singularity  of our obtained metric solutions we must be
 determine corresponding
 Ricci and Kretschmann scalars which up to second order terms become respectively \begin{equation} R=R^{\lambda}_{\lambda}=\epsilon
 \{\ddot{b_1}+2\ddot{c_1}+2(3\dot{c_1}+\dot{b_1}-\dot{a_1})/t+2(c_1-a_1)/t^2\}\end{equation} and
  \begin{equation} K=R_{\mu\nu\eta\delta}R^{\mu\nu\eta\delta}=\epsilon^2
  \{\ddot{b_1}^2+(3/2)(\ddot{c_1}+(2\dot{c_1}-\dot{a_1})/t)^2$$$$+2\dot{b_1}^2/t^2
  +24(\dot{c_1}/t+(c_1-a_1)/t^2)^2\}.\end{equation}
  Up to second order terms, one can calculate covariant conservation condition of the stress
  tensor (21) given by $\nabla_{\mu} T^{\mu}_{\nu}=0$ as
  \begin{equation}\dot{\rho}+(2/t+\epsilon(\dot{b_1}+2\dot{c_1})/2)(\rho+p)\simeq(1/3)(p_r-p_t)(2/t$$$$
  +\epsilon({\dot{c_1}}-\dot{b_1})).\end{equation}
Inserting (25), (26), (27) and (28) the above conservation
condition leads to the following form.
\begin{equation}3\dot{c_1}-\dot{a_1}-(2c_1+a_1)/t\simeq0.\end{equation}
Coefficients of linearized differential equations (15), (16) and
(17) have singularity at time $t=0$  and so it will be useful to
obtain time dependent metric solutions $(a_1,b_1,c_1)$ at
neighborhood $t\to0.$ Also we need to fix initial conditions on
the obtained solution. We will consider two different initial
conditions as $a_1(0)=b_1(0)=c_1(0)=0$ and
$a_1(\infty)=b_1(\infty)=c_1(\infty)=0$ and obtain two class of
metric solutions which are flat Minkowski at $t=0$ and
$t\to\infty$ respectively.\\
 Asymptotically behavior of the
equations (15), (16) and (17) at $t\to0$ are respectively
\begin{equation}
(8+3\omega)\dddot{c_1}+\frac{(24+11\omega)}{t}\ddot{c_1}-\frac{2\omega}{t^2}\dot{c_1}-\frac{(16+10\omega)}{t^3}c_1
$$$$+\frac{4(2+\omega)}{t}\ddot{b}_1-\frac{4(2+\omega)}{t^2}\dot{b_1}$$$$-\frac{(8+3\omega)}{t}\ddot{a_1}
-\frac{(16+6\omega)}{t^2}\dot{a_1}+\frac{(16+10\omega)}{t^3}a_1\approx0,\end{equation}
\begin{equation}(4+\omega)\ddddot{c_1}+\frac{(20+5\omega)}{t}\dddot{c_1}+\frac{(4+\omega)}{t^2}\ddot{c_1}+\frac{2(4+\omega)}{t^3}
\dot{c_1}$$$$+\frac{2(4+\omega)}{t^4}c_1+(2+\omega)\ddddot{b_1}+\frac{(12+4\omega)}{t}\dddot{b_1}
-\frac{(4+\omega)}{t}\dddot{a_1}$$$$-\frac{(4+\omega)}{t^2}\ddot{a_1}
-\frac{(8+2\omega)}{t^3}\dot{a_1}-\frac{(8+2\omega)}{t^4}a_1\approx0
\end{equation}
and
\begin{equation}(4+3\omega)\ddddot{c_1}+\frac{(40+20\omega)}{t}\dddot{c_1}+\frac{2(4+13\omega)}{t^2}\ddot{c_1}+\frac{2(5\omega-4)}{t^3}
\dot{c_1}$$$$-\frac{2(4+\omega)}{t^4}c_1+(4+\omega)\ddddot{b_1}+\frac{2(8+3\omega)}{t}\dddot{b_1}+\frac{4\omega}{t^2}\ddot{b_1}
+\frac{2\omega}{t^3}\dot{b_1}$$$$-\frac{(8+4\omega)}{t}\dddot{a_1}
-\frac{(8+12\omega)}{t^2}\ddot{a_1}+\frac{(8-6\omega)}{t^3}\dot{a_1}$$$$+\frac{(8+2\omega)}{t^4}a_1\approx0\end{equation}
where we defined dimensionless parameter $\omega$ as
\begin{equation}\omega=\frac{\beta}{\alpha}=\frac{\eta+4\xi}{\zeta-\xi}.\end{equation}
The equations (35), (36) and (37) have power-law solutions as
\begin{equation}a_1(t)\simeq  A\bigg(\frac{t}{\sqrt{\alpha}}\bigg)^{\mu},~b_1(t)\simeq B\bigg(\frac{t}{\sqrt{\alpha}}\bigg)^{\mu},~c_1(t)\simeq E
\bigg(\frac{t}{\sqrt{\alpha}}\bigg)^{\mu}\end{equation} in which
the constants $A,B,E$ and $\mu$ are related to $\omega.$ They are
obtained by inserting (39) into the equations (35), (36) and (37)
as follows.
\begin{equation}\frac{A}{B}=\{\omega^2(-3\mu^7+8\mu^6+2\mu^5-12\mu^4+13\mu^3+28\mu^2-108\mu)$$$$-\omega(14\mu^7-36\mu^6
+28\mu^4-102\mu^3+168\mu^2+104\mu)$$$$-16\mu^7+32\mu^6+32\mu^5-96\mu^4+240\mu^3
-448\mu^2+192\mu\}/\{\omega^2$$$$\times(4\mu^5-20\mu^4+90\mu^2+42\mu-140)$$$$+\omega(32\mu^5-144\mu^4+16\mu^3+276\mu^2+88\mu-144)$$$$
+(64\mu^5-256\mu^4+64\mu^3+320\mu^2 -64\mu+128)\}\end{equation}
\begin{equation}
\frac{E}{B}=-\{\omega^2(7\mu^6+365\mu^5-1923\mu^4+3107\mu^3-1536\mu^2-36\mu)$$$$+\omega(36\mu^6+606\mu^5
-3526\mu^4+5738\mu^3-2704\mu^2-248\mu)+48\mu^6$$$$-240\mu^5+624\mu^4-944\mu^3+704\mu^2-320\mu\}/\{\omega^2(4\mu^5-20\mu^4$$$$+90\mu^2+42\mu-140)
+\omega(32\mu^5-144\mu^4+16\mu^3+276\mu^2+88\mu$$$$-144)+64\mu^5-256\mu^4+64\mu^3+320\mu^2-64\mu+128\}\end{equation}
and
\begin{equation}\omega^3\{
320\mu^{9}-960\mu^8-3520\mu^7+29440\mu^6-80896\mu^5+111936\mu^4$$$$-92352\mu^3+52800\mu^2-10880\mu-2048\}+
\omega^2\{33\mu^{9}+1073\mu^8$$$$-5104\mu^7+5658\mu^6+4611\mu^5
-19253\mu^4+28128\mu^3-14094\mu^2$$$$-5308\mu+3648\}+\omega\{
464\mu^{9}+1544\mu^8
-4984\mu^7-63368\mu^6$$$$+304688\mu^5-540776\mu^4+441036\mu^3-149624\mu^2+12944\mu
$$$$+3456\}+216\mu^{9}+3158\mu^8-11818\mu^7-27952\mu^6+182054\mu^5$$$$-336406\mu^4
+298530\mu^3-115962\mu^2+236\mu+8672=0\end{equation} Inserting
(39) the conservation equation (34) leads to the following
condition.
\begin{equation}(3\mu-2)E-(1+\mu)A=0.\end{equation} Inserting (40) and (41), the latter equation can be rewritten as
\begin{equation}\omega^2
(3\mu^7-26\mu^6-1091\mu^5+6509\mu^4-13168\mu^3+10781\mu^2$$$$-2884\mu+36)+
\omega(14\mu^7-130\mu^6-1782\mu^5+11818\mu^4-24340\mu^3$$$$+19654\mu^2-4392\mu-392)+
16\mu^7-160\mu^6$$$$+752\mu^5-2288\mu^4+3936\mu^3-3792\mu^2+2624\mu-832=0.\end{equation}
One can solve (42) and (44) simultaneously to obtain numerical
values of the parameters $\mu$ and $\omega.$  We plotted diagrams
of the equations (42) and (44) at figure 1 where crossing points
determine numerical values of the parameters $\mu$ and $\omega.$
There are 12 crossing points called as
$P_i\equiv(\mu_i,\omega_i);~i=1,2,3\cdots12$ and collected at
first and second column at table 1.\\
 Inserting (12), (14) and  (39) one can obtain explicit form of the line element (13),
 the barotropic index (29) and the anisotropy index (30)
 respectively as
 \begin{equation} ds^2\approx
\bigg[1+\frac{A}{B}\bigg(\frac{t}{\sqrt{\alpha}}\bigg)^{\mu}\bigg]dt^2+\bigg[1+\bigg(\frac{t}{\sqrt{\alpha}}\bigg)^{\mu}\bigg]dr^2
$$$$+t^2\bigg[1+\frac{E}{B}\bigg(\frac{t}{\sqrt{\alpha}}\bigg)^{\mu}\bigg]
(d\theta^2+\sin^2\theta d\varphi^2)\end{equation}
\begin{equation}\Delta(t)\approx\frac{(2\mu^2+3\mu+1)\frac{E}{B}-(1+2\mu)\frac{A}{B}-\mu}{2(1+\mu)\frac{E}{B}-2\frac{A}{B}+2\mu}
\end{equation}
and
\begin{equation}\gamma(t)\approx-\frac{1}{3}\bigg[\frac{\frac{A}{B}-(3+2\mu)\frac{E}{B}-\mu}{\frac{A}{B}-(1+\mu)\frac{E}{B}-\mu}\bigg]\end{equation}
where we set
\begin{equation}\epsilon=\frac{1}{B}.\end{equation} Inserting numerical values of the parameters $\mu_i,\omega_i$ given in the table 1 one can obtain
numerical values of the coefficients $\frac{A}{B},$ $\frac{E}{B},$
$\Delta(t)$ and $\gamma(t)$ for all 12 points $P_i.$ They are
collected also at the figure 1.\\
Inserting (39) we can rewrite the equations (25), (26), (27),
(28), (32) and (33) as follows.
\begin{equation}\frac{\rho(t)}{\rho(\sqrt{\alpha})}\approx\bigg(\frac{t}{\sqrt{\alpha}}\bigg)^{\mu-2}=\frac{p_r(t)}{p_r(\sqrt{\alpha})}
=\frac{p_t(t)}{p_t(\sqrt{\alpha})}=\frac{p(t)}{p(\sqrt{\alpha})}$$$$=
\frac{R^{\lambda}_{\lambda}(t)}{R^{\lambda}_{\lambda}(\sqrt{\alpha})}=\sqrt{\frac{K(t)}{K(\sqrt{\alpha})}}
\end{equation}
where we defined
\begin{equation}\rho^*=8\pi\alpha\rho(\sqrt{\alpha})=\bigg[1-\frac{A}{B}+(1+\mu)\frac{E}{B}\bigg]\end{equation}
\begin{equation}p_r^*=8\pi\alpha p_r(\sqrt{\alpha})=(1+\mu)\bigg[\frac{A}{B}-(1+\mu)\frac{E}{B}\bigg]\end{equation}
\begin{equation}p_t^*=8\pi\alpha p_t(\sqrt{\alpha})=\frac{\mu}{2}\bigg[\frac{A}{B}-(1+\mu)\frac{E}{B}-\mu\bigg]\end{equation}
\begin{equation}p^*=8\pi\alpha p(\sqrt{\alpha})=\frac{\mu}{3}\bigg[\frac{A}{B}-(3+2\mu)\frac{E}{B}-\mu\bigg]\end{equation}
\begin{equation}R^*=\alpha R^{\lambda}_{\lambda}(\sqrt{\alpha})=(1+\mu)\bigg[\mu-2\frac{A}{B}+2(1+\mu)\frac{E}{B}\bigg]\end{equation}
\begin{equation}K^*=\alpha^2K(\sqrt{\alpha})=\bigg[\mu^2(\mu^2-2\mu+3)+25\bigg(\frac{A}{B}\bigg)^2$$$$+(1+\mu)^2[24+3\mu^2/2]\bigg(
\frac{E}{B}\bigg)^2-2(24+\mu)(1+\mu)\bigg(\frac{E}{B}\bigg)\bigg(\frac{A}{B}\bigg)
\bigg].\end{equation} Numerical values of the parameters (50),
(51), (52), (53), (54) and (55) are calculated at the particular
points $P_i;~ i=1,2,3,\cdots12$ and are collected in table 2. In
the next section we proceed  to obtain the times where internal
event and apparent horizons of our obtained metric solutions are
formed.

\section{Horizons formation and trapped surfaces}

Here we assume the existence of an apparent horizon, which is
defined as the outer boundary of a connected component of the
trapped region. The important feature of the apparent horizon is
that, if the space time is strongly asymptotically predictable and
the null convergence condition holds, the presence of the apparent
horizon implies the desistance of an event horizon outside or
coinciding with it. If the connected component of the trapped
region has the structure of a manifold with boundaries, then the
apparent horizon is an outer marginally trapped surface with
vanishing expansion [19]. Along a future directed outgoing null
geodesic, the relation [20]
\begin{equation}\frac{dR}{dt}=\partial_tR+\partial_rR\frac{dr}{dt}$$$$=e^{\nu}\bigg(e^{-\lambda}\partial_rR\pm\sqrt{-1+\frac{2m}{R}+e^{-2\lambda}
(\partial_rR)^2}\bigg)\end{equation}
 for line element (7) is satisfied, where $+ (-)$ corresponds to
 expanding (collapsing) phase and \begin{equation} m(t,r)=\frac{R}{2}\big(1-e^{-2\lambda}(\partial_rR)^2+e^{-2\nu}(\partial_tR)^2\big)\end{equation}
is the Misner-Sharp mass function [21]. Assuming $\partial_rR>0$
the equation (56) shows that in the expanding phase there is no
apparent horizon but in the collapsing phase on a hypersurface of
constant $t$ the two-sphere $R=2m$ is an apparent horizon. The
region $R<2m$ is trapped surfaces where $\frac{dR}{dt}<0$, while
the region $R>2m$ is not trapped in which $\frac{dR}{dt}>0.$
Singularities which can be appeared in spherically collapse are
called `shell-crossing` if $\partial_rR=0$ with $R>0$ and
`shell-focusing` if $R=0.$ In the local frame, the space time may
to be has `central-singularity` and `non-central singularity`
which is characterized by $r=0$ and $r>0$ respectively.\\
  In the metric solution (45) time dependent 2-sphere radiuses are
\begin{equation}R(t,r)=t\sqrt{1+\frac{E}{B}\bigg(\frac{t}{\sqrt{\alpha}}\bigg)^{\mu}}\end{equation}
in which $\partial_rR=0$ and so all hypersurfaces $R>0$ are
shell-crossing type of space time singularity. Also the particular
hypersurfaces
\begin{equation}
t_1=0,~~~~~t_2=\sqrt{\alpha}\left(-\frac{B}{E}\right)^{\frac{1}{\mu}}
\end{equation} which are solutions of the equation $R(t)=0$ are shell-focusing singularity of our metric
solution. Trapped surfaces are determined by equating
$\frac{dR}{dt}\leq0$ as
\begin{equation}R(t)=constant,~~~~0\leq\bigg(\frac{t}{\sqrt{\alpha}}\bigg)\leq
\frac{T_{AH}}{\sqrt{\alpha}}
\end{equation}
in
which\begin{equation}\frac{T_{AH}}{\sqrt{\alpha}}=\bigg[\bigg(\frac{-2}{2+\mu}\bigg)\frac{B}{E}\bigg]^{\frac{1}{\mu}}\end{equation}
is apparent horizon formation time obtained from $\frac{dR}{dt}=0$
with corresponding radius
\begin{equation} \frac{R(T_{AH})}{\sqrt{\alpha}}=\bigg[\bigg(\frac{-2}{2+\mu}\bigg)\frac{B}{E}\bigg]
^{\frac{1}{\mu}}\sqrt{\frac{\mu}{2+\mu}}.\end{equation}
 Event horizon formation time $T_{EH}$ is determined by solving the
equation $g_{tt}(t)=0$ given by (45) as
\begin{equation}\frac{T_{EH}}{\sqrt{\alpha}}=\bigg(-\frac{A}{B}\bigg)^{-\frac{1}{\mu}}\end{equation} with corresponding 2-sphere radiuses
\begin{equation}\frac{R(T_{EH})}{\sqrt{\alpha}}=\bigg(-\frac{B}{A}\bigg)^{\frac{1}{\mu}}\sqrt{1-\frac{E}{A}}.
\end{equation}
 Black holes are formed with total Misner-Sharp mass $M=m(T_{AH})$ and radius $R(T_{AH})$ if
\begin{equation}R_{EH}\geq R_{AH}\end{equation}
which by inserting (62) and (64) leads to
\begin{equation}\sigma=\frac{E}{A}+\bigg(\frac{2}{2+\mu}\bigg)^{2\mu}\bigg(\frac{\mu}{2+\mu}\bigg)-1\geq0.\end{equation}
If our metric solutions given by $P_i;~ i=1,2,3,\cdots12$ at table
1 can not satisfy the above condition then they will be describe
formation of  naked singularity at end of collapse. Numerical
values of the quantity $\sigma$ are given in the table 3 for all
our metric solutions. The choice $P_3$ describes a collapsing
cloud which reaches finally to a covered singularity by standing
apparent horizon and forming event horizon after than that with
larger radius. Other cases $P_i;i\neq3$ given in the table 3 may
show naked singularity as end of the collapse. But since the
absence of an apparent horizon does not necessary implies the
absence of an event horizon, thus we seek naked singularity
formation or otherwise also by evaluating time dependence of
radial null geodesic expansion parameter as follows. It will be
help us to determine nakedness of singularity of other metric
solutions denoted with $P_i;i\neq3$ via studding of trapped
surfaces in the following section. Our results are collected in
the table 4.

\section{Conditions on radial null geodesic expansion}

Consider a congruence of outgoing radial null geodesics having the
tangent vector $(V^t(\tau), V^r(\tau),0,0),$ where
\begin{equation}V^t(\tau)=\frac{dt}{d\tau},~~~~V^r(\tau)=\frac{dr}{d\tau}\end{equation}
 and $\tau$ is an affine parameter,
along the geodesics. In terms of these two vector fields the
geodesic equation can be written as
\begin{equation}\frac{d\ln V^t}{d\tau}=-
\bigg\{\bigg(\frac{\mu}{\sqrt{\alpha}}\bigg)\bigg(\frac{A}{B}\bigg)\frac{\big(\frac{t}{\sqrt{\alpha}}\big)^{\mu-1}}{1+\frac{A}{B}
\big(\frac{t}{\sqrt{\alpha} }\big)^\mu}\bigg\}V^t\end{equation}
and
\begin{equation}\frac{d\ln V^r}{d\tau}=-\bigg\{
\bigg(\frac{\mu}{\sqrt{\alpha}}\bigg)\frac{\big(\frac{t}{\sqrt{\alpha}}\big)^{\mu-1}}{1+\big(\frac{t}{\sqrt{\alpha}}\big)^\mu}\bigg\}V^t.\end{equation}
The above equations show that
\begin{equation}\frac{\partial V^t}{\partial r}=0,~~~~\frac{\partial V^r}{\partial r}=0.\end{equation}
 The geodesic expansion parameter
is defined by $\Theta=\nabla_\mu V^\mu=\partial_\mu
V^\mu+\Gamma_{\mu\nu}^\mu V^\nu$ which by applying (70) and our
metric solution (45) reduces to the following
form.\begin{equation}\Theta=\frac{\partial V^t}{\partial
t}+\frac{\Omega(t)V^t}{2\sqrt{\alpha}}\end{equation} where
\begin{equation}\Omega(t)=\big(\frac{t}{\sqrt{\alpha}}\big)^{-1}
\bigg\{2+\big[3\big(1+\frac{A}{B}\big)+(2+\mu)
\frac{E}{B}\big]\big(\frac{t}{\sqrt{\alpha}}\big)^{\mu}$$$$+\big[4\frac{A}{B}+(4+\mu)\frac{E}{A}
+(2+\mu)\frac{AE}{B^2}\big]\big(\frac{t}{\sqrt{\alpha}}
\big)^{2\mu}+\frac{EA}{B^2}(2+\mu)$$$$\times
\big(\frac{t}{\sqrt{\alpha}}\big)^{3\mu}\bigg\}/\bigg\{\big[1+\big(\frac{t}{\sqrt{\alpha}}\big)^\mu\big]
\big[1+\frac{A}{B}
\big(\frac{t}{\sqrt{\alpha}}\big)^\mu\big]\big[1+\frac{E}{B}\big(\frac{t}{\sqrt{\alpha}}\big)^\mu\big]\bigg\}.\end{equation}
First term of RHS of the equation (71) can be rewritten as
\begin{equation}\frac{d V^t}{d \tau}=\frac{\partial
V^t}{\partial t}\frac{
\partial t}{\partial \tau}.\end{equation} Inserting (67) into the above relation
one can result
\begin{equation}\frac{\partial V^t}{\partial t}=\frac{d\ln V^t}{d\tau}.\end{equation}
Inserting (68) and (74) we can rewrite the geodesics expansion
parameter (71) such as follows.
\begin{equation}\Theta=\frac{\Xi(t)}{2\sqrt{\alpha}}V^t\end{equation} where we defined
 \begin{equation}\Xi(t)=\big(\frac{t}{\sqrt{\alpha}}\big)^{-1}
\bigg\{2+\big[3+(3-2\mu)\frac{A}{B}+(2+\mu)
\frac{E}{B}\big]\big(\frac{t}{\sqrt{\alpha}}\big)^{\mu}$$$$+\big[2(2-\mu)\frac{A}{B}+(4+\mu)\frac{E}{A}
+(2-\mu)\frac{AE}{B^2}\big]\big(\frac{t}{\sqrt{\alpha}}
\big)^{2\mu}+\frac{EA}{B^2}(2-\mu)$$$$\times\big(\frac{t}{\sqrt{\alpha}}\big)^{3\mu}\bigg\}/\bigg\{\big[1+\big(\frac{t}{\sqrt{\alpha}}
\big)^\mu\big]\big[1+\frac{A}{B}
\big(\frac{t}{\sqrt{\alpha}}\big)^\mu\big]\big[1+\frac{E}{B}\big(\frac{t}{\sqrt{\alpha}}\big)^\mu\big]\bigg\}.\end{equation}
Radial ingoing (-) and outgoing (+) null geodesics of the metric
solutions (45) are obtained by setting $ds=0=d\theta=d\varphi$ as
\begin{equation}\frac{V^t}{V^r}=\frac{dt}{dr}=\pm\sqrt{-\bigg(\frac{1+\big(\frac{t}{\sqrt{\alpha}}\big)^\mu}{1+\frac{A}{B}
\big(\frac{t}{\sqrt{\alpha}}\big)^{\mu}}\bigg)}\end{equation}
where we use (67). Eliminating  `t` and $\tau$ parameters between
(68), (69) and (77) we obtain
\begin{equation} d\bigg(\frac{1}{(V^t)^2}-\frac{A}{B}\frac{1}{(V^r)^2}\bigg)=0\end{equation}
which leads to the following relation.
\begin{equation} \frac{1}{(V^t)^2}-\frac{A}{B}\frac{1}{(V^r)^2}=\Upsilon=constant.\end{equation} Using the above relation we can choose
\begin{equation}V^t(\tau)=\frac{1}{\sqrt{\Upsilon}}\frac{1}{\cosh\tau},~~~V^r(\tau)=\sqrt{\frac{A}{B\Upsilon}}\frac{1}{\sinh\tau}\end{equation}
 where \begin{equation} \tanh(\tau)=\pm\sqrt{-\bigg(\frac{\frac{A}{B}+\frac{A}{B}\big(\frac{t}{\sqrt{\alpha}}\big)^\mu}{1+\frac{A}{B}
\big(\frac{t}{\sqrt{\alpha}}\big)^{\mu}}\bigg)}.\end{equation}
Using (80) and (81) we obtain\begin{equation}
V^t(t)=\frac{1}{\sqrt{\Upsilon}}\sqrt{\frac{1+\frac{A}{B}+2\frac{A}{B}\big(\frac{t}{\sqrt{\alpha}}\big)^\mu}{1+\frac{A}{B}\big(\frac{t}{
\sqrt{\alpha}}\big)^\mu}}\end{equation} and
\begin{equation}V^r(t)=\frac{\pm1}{\sqrt{\Upsilon}}\sqrt{\frac{1+\frac{A}{B}+2\frac{A}{B}\big(\frac{t}{\sqrt{\alpha}}\big)^\mu}
{-\big(1+\big(\frac{t}{
\sqrt{\alpha}}\big)^\mu\big)}}.\end{equation} Inserting (82), the
geodesics expansion parameter (75) can be rewritten as
\begin{equation}\Theta^{*}(t)=2\sqrt{\alpha\Sigma}\Theta(t)=\big(\frac{t}{\sqrt{\alpha}}\big)^{-1}\sqrt{\frac{1+\frac{A}{B}+2\frac{A}{B}
\big(\frac{t}{\sqrt{\alpha}}\big)^\mu}{1+\frac{A}{B}\big(\frac{t}{
\sqrt{\alpha}}\big)^\mu}}$$$$\times
\bigg\{2+\big[3+(3-2\mu)\frac{A}{B}+(2+\mu)
\frac{E}{B}\big]\big(\frac{t}{\sqrt{\alpha}}\big)^{\mu}+\big[2(2-\mu)\frac{A}{B}$$$$+(4+\mu)\frac{E}{B}
+(2-\mu)\frac{AE}{B^2}\big]\big(\frac{t}{\sqrt{\alpha}}
\big)^{2\mu}+\frac{EA}{B^2}(2-\mu)\big(\frac{t}{\sqrt{\alpha}}\big)^{3\mu}\bigg\}/$$$$\bigg\{\big[1+\big(\frac{t}{\sqrt{\alpha}}
\big)^\mu\big]\big[1+\frac{A}{B}
\big(\frac{t}{\sqrt{\alpha}}\big)^\mu\big]\big[1+\frac{E}{B}\big(\frac{t}{\sqrt{\alpha}}\big)^\mu\big]\bigg\}\end{equation}where
when $\Theta>0$ the singularity become visible and so cosmic
censorship conjecture is violated but not for $\Theta\leq0.$ In
the latter case trapped surfaces appear and their boundary surface
region called as apparent horizon is determined by $\Theta=0.$
Namely all trapped surfaces satisfy $\Theta<0.$ These surfaces are
closed orientable smooth two dimensional space-like surfaces such
that both families of ingoing and outgoing null geodesics
orthogonal to them necessarily converge. The singularity is called
naked if there exist a family of future directed non-space-like
geodesics reaching faraway observers in space-time and terminating
at the singularity in the past with a definite tangent. If such
family of curves do not exist and the event horizon forms earlier
than the singularity covering it, a black hole is formed. We
should point that the absence of an apparent horizon dose not
necessarily implies the absence of an event horizon. Some of our
obtained metric solutions exhibit with event horizon formation
while an apparent horizon can not to form (see table 3). Inserting
numerical values of the parameters $\{\mu_i,(A/B)_i,(E/B)_i\}$
from the table 1 we obtain exact form of the null geodesics
expansion parameter $\Theta^*(t)_i$ for $i=1,2,3,\cdots 12$
respectively as follows.
\begin{equation} \Theta^*_1(t)=\frac{1}{T}\sqrt{\frac{2.45+2.9T^{2.433}}{1+1.45T^{2.433}}}$$$$\times\frac{(2+4.812T^{2.433}+4.660T^{4.866}-0.640T^{7.299})}{
(1+T^{2.433}) (1+1.45T^{2.433})(1+1.019T^{2.433})}\end{equation}

\begin{equation} \Theta^*_2(t)=\frac{1}{T}\sqrt{\frac{1.029+0.058T^{2.057}}{1+0.029T^{2.057}}}$$$$\times
\frac{(2+2.907T^{2.057}-0.094T^{4.114}+0.00003T^{6.171})}{
(1+T^{2.057}) (1+0.029T^{2.057})(1-0.015T^{2.057})}\end{equation}

\begin{equation} \Theta^*_3(t)=\frac{1}{T}\sqrt{\frac{0.826-0.348T^{1.888}}{1-0.174T^{1.888}}}$$$$\times
\frac{(2+2.404T^{1.888}-1.142T^{3.776}+0.004T^{5.664})}{
(1+T^{1.888}) (1-0.174T^{1.888})(1-0.188T^{1.888})}\end{equation}

\begin{equation} \Theta^*_4(t)=\frac{1}{T}\sqrt{\frac{2.633+3.266T^{1.616}}{1+1.633T^{1.616}}}$$$$\times
\frac{(2+8.291T^{1.616}+11.043T^{3.232}+0.983T^{4.848})}{
(1+T^{1.616}) (1+1.633T^{1.616})(1+1.568T^{1.616})}\end{equation}

\begin{equation} \Theta^*_5(t)=\frac{1}{T}\sqrt{\frac{0.999-0.002T^{0.644}}{1-0.001T^{0.644}}}$$$$\times
\frac{(2+3.778T^{0.644}+1.367T^{1.288}-0.0004T^{1.932})}{
(1+T^{0.644}) (1-0.001T^{0.644})(1+0.295T^{0.644})}\end{equation}

\begin{equation} \Theta^*_6(t)=\frac{1}{T}\sqrt{\frac{2.478+2.956T^{0.521}}{1+1.478T^{0.521}}}$$$$\times
\frac{(2-6.162T^{0.521}-27.701T^{1.042}-10.453T^{1.563})}{
(1+T^{0.521}) (1+1.478T^{0.521})(1-4.782T^{0.521})}\end{equation}

\begin{equation} \Theta^*_7(t)=\frac{1}{T}\sqrt{\frac{0.696-0.608T^{0.508}}{1-0.304T^{0.508}}}$$$$\times
\frac{(2+5.121T^{0.508}+3.496T^{1.016}-0.493T^{1.524})}{
(1+T^{0.508}) (1-0.304T^{0.508})(1+1.086T^{0.508})}\end{equation}

\begin{equation} \Theta^*_8(t)=\frac{1}{T}\sqrt{\frac{0.972-0.056T^{-0.041}}{1-0.028T^{-0.041}}}$$$$\times
\frac{(2+2.896T^{-0.041}-0.149T^{-0.082}+0.0005T^{-0.123})}{
(1+T^{-0.041})
(1-0.028T^{-0.041})(1-0.009T^{-0.041})}\end{equation}

\begin{equation} \Theta^*_9(t)=\frac{1}{T}\sqrt{\frac{1.389+0.778T^{-0.201}}{1+0.389T^{-0.201}}}$$$$\times
\frac{(2+3.115T^{-0.201}-1.416T^{-0.402}-0.575T^{-0.603})}{
(1+T^{-0.201})
(1+0.389T^{-0.201})(1-0.672T^{-0.201})}\end{equation}

\begin{equation} \Theta^*_{10}(t)=\frac{1}{T}\sqrt{\frac{0.956-0.088T^{-0.270}}{1-0.044T^{-0.270}}}$$$$\times
\frac{(2+2.763T^{-0.270}-0.370T^{-0.540}+0.005T^{-0.810})}{
(1+T^{-0.270})
(1-0.044T^{-0.270})(1-0.047T^{-0.270})}\end{equation}

\begin{equation} \Theta^*_{11}(t)=\frac{1}{T}\sqrt{\frac{4.467+6.934T^{-3.284}}{1+3.467T^{-3.284}}}$$$$\times
\frac{(2+37.892T^{-3.284}+11.151T^{-6.568}-24.530T^{-9.852})}{
(1+T^{-3.284})
(1+3.467T^{-3.284})(1-1.339T^{-3.284})}\end{equation}

\begin{equation} \Theta^*_{12}(t)=\frac{1}{T}\sqrt{\frac{8.146+18.292T^{-6.206}}{9.146T^{-6.206}-1}}$$$$\times
\frac{(2-127.892T^{-6.206}+37.865T^{-12.412}+182.602T^{-18.616})}{
(1+T^{-6.206})
(1-9.146T^{-6.206})(1-2.433T^{-6.206})}\end{equation}

where we defined
\begin{equation}T=\frac{t}{\sqrt{\alpha}}.\end{equation}
Diagrams of the null geodesics expansion parameter
$\Theta^*_i(t);~i=1,2,3,\cdots12$ are plotted at figures
$1,2,3,\cdots12$ respectively against $t.$ and their descriptions
are given as follows.
\begin{description}
    \item[Figure 2.]  Diagram crosses horizontal axis at $T_1=2.383$ and
$T\to\infty$ (thee singularity). Trapped surfaces is characterized
for times $2.383<T<\infty$ and so the singularity is located
inside of trapped surfaces. Thus final state of the collapse is
covered singularity as big black hole.
    \item[Figure 3.] Diagram crosses the horizontal axis at
$T_1=5.818,$ $T_2=7.665$ and $T_3\to\infty$ (the singularity).
$\Theta_2^*$ takes negative values for $5.818<T<7.665$ where
trapped surfaces are happened and positive values for
$7.665<T<\infty.$ So the singularity is located outside of trapped
surfaces. Thus final state of collapse is naked singularity.
    \item[Figure 4.] Here  $\Theta_3^*$ takes positive values for
finite times $0<T<1.707$ (apparent horizon formation time) and it
takes negative values (trapped surfaces) for times $T>2.536$
(event horizon formation time) and so the trapped surfaces located
inside of trapped surfaces. Thus final state of the collapse
reaches to a covered singularity (black hole).
    \item[Figure 5.] Here $\Theta_4^*>0$ and so there is not
trapped surfaces and so final state of the collapse reaches to a
naked singularity.
    \item[Figure 6.]  Here $\Theta_5^*>0$ and so there is not
trapped surfaces and so final state of the collapse reaches to a
naked singularity.
    \item[Figure 7.] Here $\Theta_6^*>0$ for $0<T<0.036$ and
$T\geq0.05$ while it takes negative values for $0.036<T<0.05.$
Singularity is $T=0$ and so it become invisible from distant
observers.
    \item[Figure 8.] Here $\Theta_7^*>0$ for finite times
$0<T<1.334$ and so there is not trapped surfaces and thus the
singularity $T=0$ become visible from view of distant observers.
    \item[Figure 9.]  Here $\Theta_8^*>0$ and so there is not
trapped surfaces and so final state of the collapse reaches to a
naked singularity.
    \item[Figure 10.]  Here $\Theta_9^*>0$ for $0<T<0.058$ and
$T\geq0.139$ while it takes negative values for $0.058<T<0.139.$
Singularity is $T=0$ and so it become invisible from distant
observers.
    \item[Figure 11.] Here $\Theta_6^*>0$ for $T>0.00025$ and so
singularity  $T=0$ is located inside of trapped surfaces. Thus
singularity become invisible from point of view of distant
observers.
    \item[Figure 12.]  Here $\Theta_{11}^*>0$ for $0<T<0.882$ and
$T\geq1.097$ while it takes negative values for $0.882<T<1.097.$
Singularity is $T=0$ and so it become invisible from distant
observers.
\item[Figure 13.]  Here $\Theta_{12}^*>0$ for $0<T<1.065$ and
$T\geq1.148$ while it takes negative values for $1.065<T<1.148.$
Singularity is $T=0$ and so it become invisible from distant
observers.
   \end{description}
  \newpage \begin{center}    Table 1. Characteristics of metric parameters
 \end{center}
 \begin{center}
\begin{tabular}{|c|c|c|c|c|c|c|}
 \hline
  $P_i$ & $\mu_i$ & $\omega_i$ & $(A/B)_i$ & $(E/B)_i$ &  $R^*_i$   &  $K^*_i$  \\
  \hline
  $P_1$ & +2.433 &  -0.719 &  +1.450 &  +1.019 & -25.619& +210.786\\
  $P_2$ & +2.057 & +0.968 &   +0.029  &  -0.015  &  +6.392 & +13.345\\
  $P_3$ & +1.888 & -0.607 &    -0.174  &  -0.188 &  +9.583 & +14.424 \\
  $P_4$ & +1.616 & -1.648 &    +1.633  &   +1.568  & -25.774  & +199.272\\
  $P_5$ & +0.644& -0.086 &    -0.001  &  +0.295 & -0.531 &   +6.690 \\
  $P_6$ &+0.521 & +1.883 &    +1.478  & -4.782 &  +18.432 &  +1875.170  \\
  $P_7$ & +0.508 &-1.713 &   -0.304   &   +1.086  & -3.257 &  +92.697 \\
  $P_8$ & -0.041 & +3.285 &  -0.028   &   -0.009  & -0.002& +0.039 \\
  $P_9$ & -0.201 & +0.850 &   +0.389  &  -0.672   & +0.076  &  +20.787 \\
  $P_{10}$ & -0.270 & -1.986 &  -0.044 & -0.047  & -0.083 &  +0.269 \\
 $P_{11}$ & -3.284 &  -2.013 &  +3.467   &  -1.339  & +37.305 & +456.394 \\
  $P_{12}$& -6.206 &+0.645 &    -9.146   & -2.433  &  +68.973 &  +21412.731 \\
  \hline
\end{tabular}
\end{center}
\begin{center}
Table 2. Fluid characteristics of stellar cloud.
\end{center}
\begin{center}
\begin{tabular}{|c|c|c|c|c|c|c|c|}
  \hline
  $P_i$ & ${\rho_i}^*$ & ${p_i}^*_r$ & ${ p_i}^*_t$ & ${p_i}^*$ &     $\Delta_i(t)$ \\
  \hline
  $P_1$ & +3.049 &  -7.033 &  -5.452 &  -7.298 &  +1.069\\
  $P_2$ &+0.925 & +0.230 &  -2.038 &  -1.317  &  -0.616\\
  $P_3$ & +0.633 & +1.061 &   -1.435  &  -0.498 &  -1.198\\
  $P_4$ & +3.468 & -6.456 &   -3.300  &   -5.253   &  +1.081  \\
  $P_5$ & +1.486& -0.799 &   -0.364  & -0.410 &   +0.207\\
  $P_6$ &-7.755 & +13.316  &   +2.145  &  +3.525 &  +1.117 \\
  $P_7$ & +2.942 &-2.928  & -0.622   &  -0.876  &    +0.695  \\
  $P_8$ & +1.037 & -0.036&   -0.0001   &   +0.0002  & -8.917\\
  $P_9$ &+0.074 & +0.740&    -0.113 & -0.157  &   +0.157\\
  $P_{10}$ & +1.010 & -0.007 &   -0.035 &  -0.031  &  -0.527\\
 $P_{11}$ &+0.590 &  -0.936 &  -6.065  &  -2.162  &   -0.753  \\
  $P_{12}$& +22.814 &+113.561 &   +48.430   &  +53.458  &  -7.776\\
  \hline
\end{tabular}
\end{center}
\newpage
\begin{center} Table 3. Time and radius of formed event and apparent
horizons
\end{center}\begin{center}\begin{tabular}{|c|c|c|c|c|}
  \hline
  $P_i$ & $T_{EH}/\sqrt{\alpha}$ & $T_{AH}/\sqrt{\alpha}$ & $R_{EH}/\sqrt{\alpha}$ & $R_{AH}/\sqrt{\alpha}$ \\
  \hline
  $P_1$ & - & - &  - & - \\
  $P_2$ & - & +5.453 &  - &+3.889 \\
  $P_3$ & +2.526 & +1.707&  +2.752 & +1.188  \\
  $P_4$ & - & - &  - &-\\
  $P_5$ &  36692.907& - & +38236.634 & - \\
  $P_6$& - & +0.032 & - & +0.015 \\
  $P_7$ & +10.427 & - &+3.054 &- \\
  $P_8$& $1.372\times10^{-38}$ & - & $1.342\times10^{-38}$ &-  \\
  $P_9$ & - & 0.082 & - & - \\
  $P_{10}$ & $9.473\times10^{-6}$ & $7.099\times10^{-6}$ & $+9.677\times10^{-6}$ &- \\
 $P_{11}$ & - & -&  - & -\\
  $P_{12}$ & +1.429& - &  +2.647 &- \\
  \hline
\end{tabular}
\end{center}

\begin{center} Table 4. Nakedness, trapped
surfaces and phase of fluid
\end{center}\begin{center}\begin{tabular}{|c|c|c|c|c|}
  \hline
  $P_i$ & $\gamma_i(t)$ & Trapped S. & Phase of fluid & Nakedness \\
  \hline
  $P_1$ & -0.669 & Full &  Domain walls & Covered \\
  $P_2$ & -0.323 & Quasi &  Cosmic sting  &Naked   \\
  $P_3$ & -0.174 & Full &  Dark matter & Covered  \\
  $P_4$ & -0.796 & No &  Domain walls &Naked \\
  $P_5$ &  -0.563& No & Domain walls & Naked \\
  $P_6$&-0.822 & Quasi & Anti-matter & Covered \\
  $P_7$ & -0.704& No &Domain walls &Naked \\
  $P_8$&+1.092 & No & Stiff matter &Naked  \\
  $P_9$ & -0.691 &Quasi & Domain walls & Covered \\
  $P_{10}$ & -0.438 & Quasi & Quasi-cosmic string &Covered \\
 $P_{11}$ & -0.178& Quasi&  Dark matter & Covered\\
  $P_{12}$ & -0.552& Quasi &  Quasi-domain walls &Covered \\
  \hline
\end{tabular}
\end{center}

\begin{center} Table 5. Numerical values of the black hole formation condition $\sigma_i\geq0$
\end{center}\begin{center}\begin{tabular}{|c|c|}
  \hline
  $P_i$  & $\sigma_i$\\
  \hline
  $P_1$  &-0.286\\
  $P_2$ &-1.490 \\
  $P_3$ &+0.120 \\
  $P_4$ &+0.026 \\
  $P_5$ &-295.830 \\
  $P_6$&-4.073 \\
  $P_7$ &-4.411 \\
  $P_8$& -0.700  \\
  $P_9$ &-2.835 \\
  $P_{10}$ &-0.076 \\
 $P_{11}$ &-\\
  $P_{12}$ &- \\
  \hline
\end{tabular}
\end{center}

\section{Concluding remarks }

We apply an alternative higher order derivative gravity model to
obtain time-dependent
 internal metric solution of an anisotropic spherically symmetric collapsing cloud
. We obtain class of solutions (12 different kinds) where
geometrical source treats as domain walls (6 kinds), cosmic string
(2 kinds), dark matter (2 kinds), stiff matter (1 kind) and
anti-matter  ( negative energy density) with 1 kinds. In summary 7
kinds of our solutions reach to compact object with covered
singularity (the black hole) but 5 solutions reach to naked
singular metric at end of the collapse and hence cosmic censorship
conjecture maintain valid for 7 kinds of our 12 metric solutions
only.

\begin{figure}
\includegraphics[width=7cm]{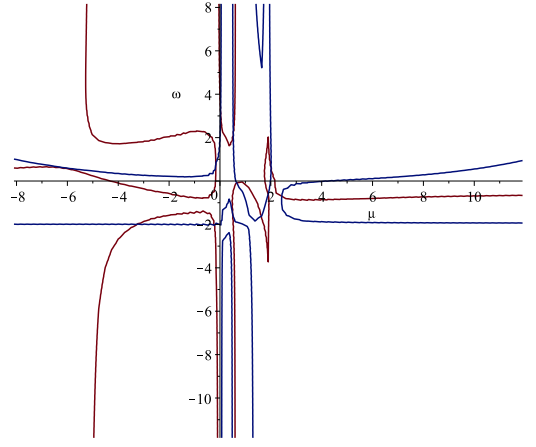}
\caption{\label{fig:epsart}  Diagram of the equations (42) and
(44) are plotted against $(\mu,\omega)$. Numerical values of
crossing points $P_i\equiv(\mu_i,\omega_i)$ with
$i=1,2,3,4,\cdots12$ are given in table 1.  }
\end{figure}
\begin{figure}
\includegraphics[width=7cm]{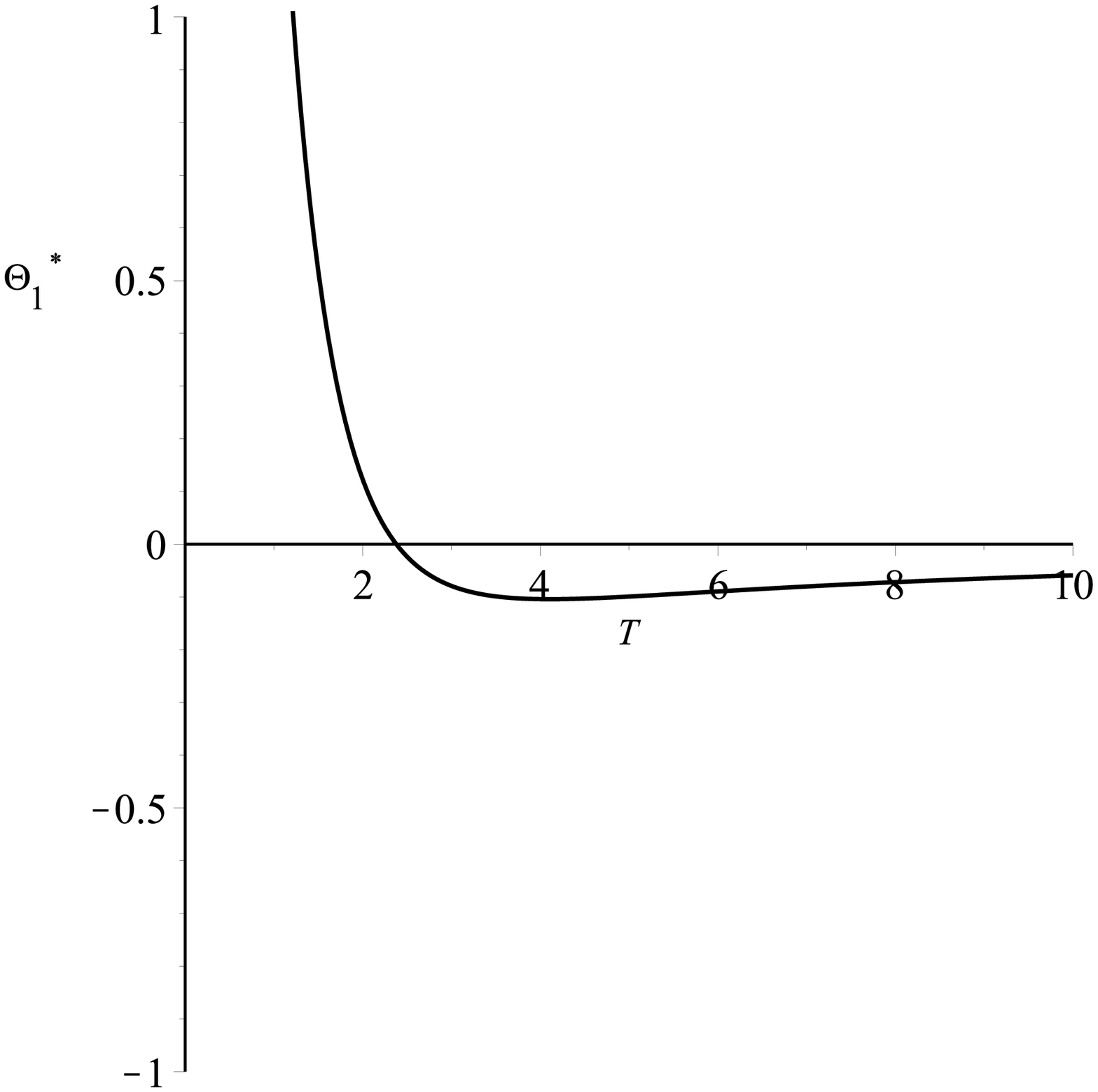}
\caption{\label{fig:epsart}   Diagram of the equation (85) is
plotted against $T.$ }
\end{figure}
\begin{figure}
\includegraphics[width=7cm]{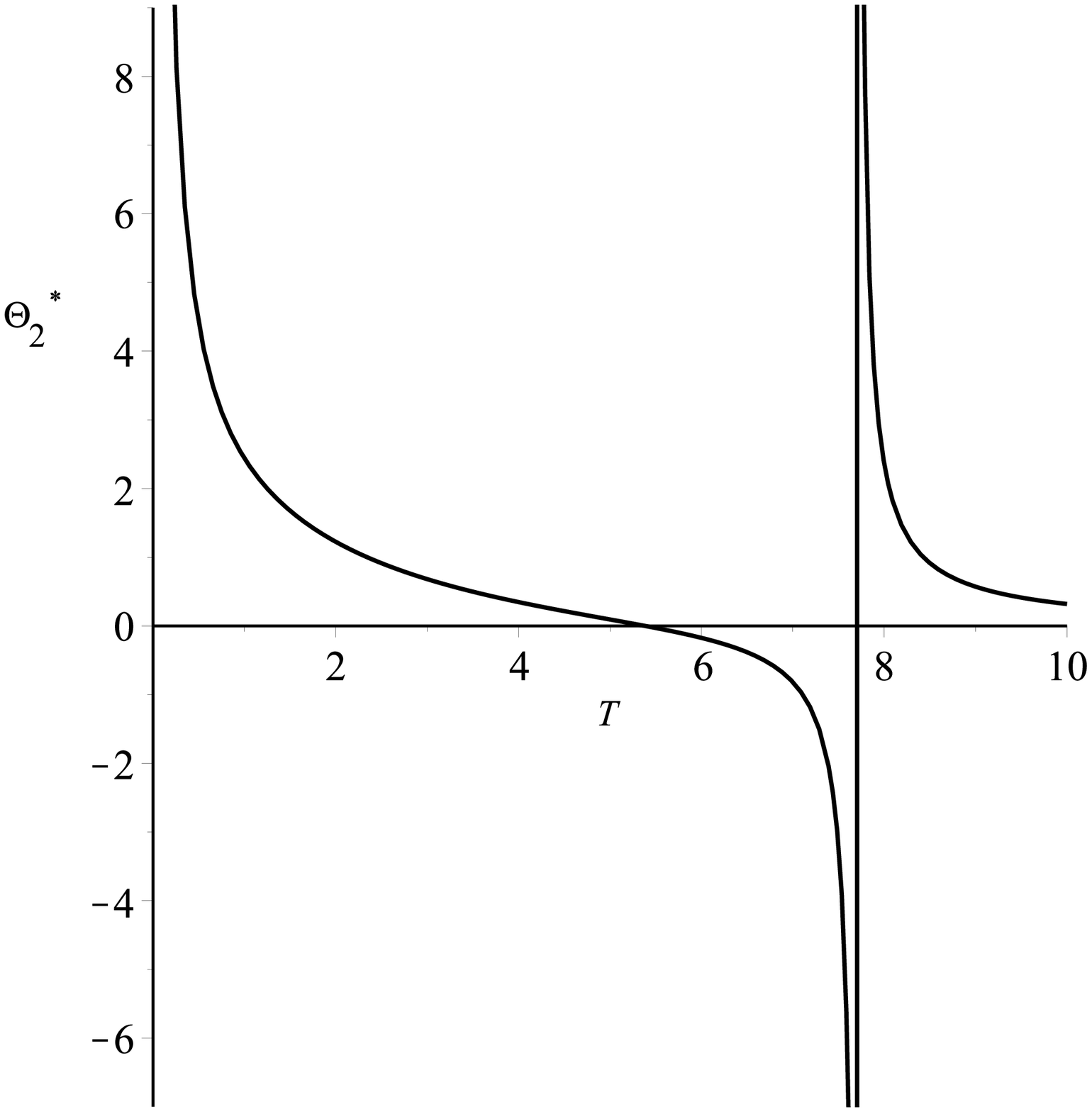}
\caption{\label{fig:epsart}  Diagram of the equation (86) is
plotted against $T.$   }
\end{figure}
\begin{figure}
\includegraphics[width=7cm]{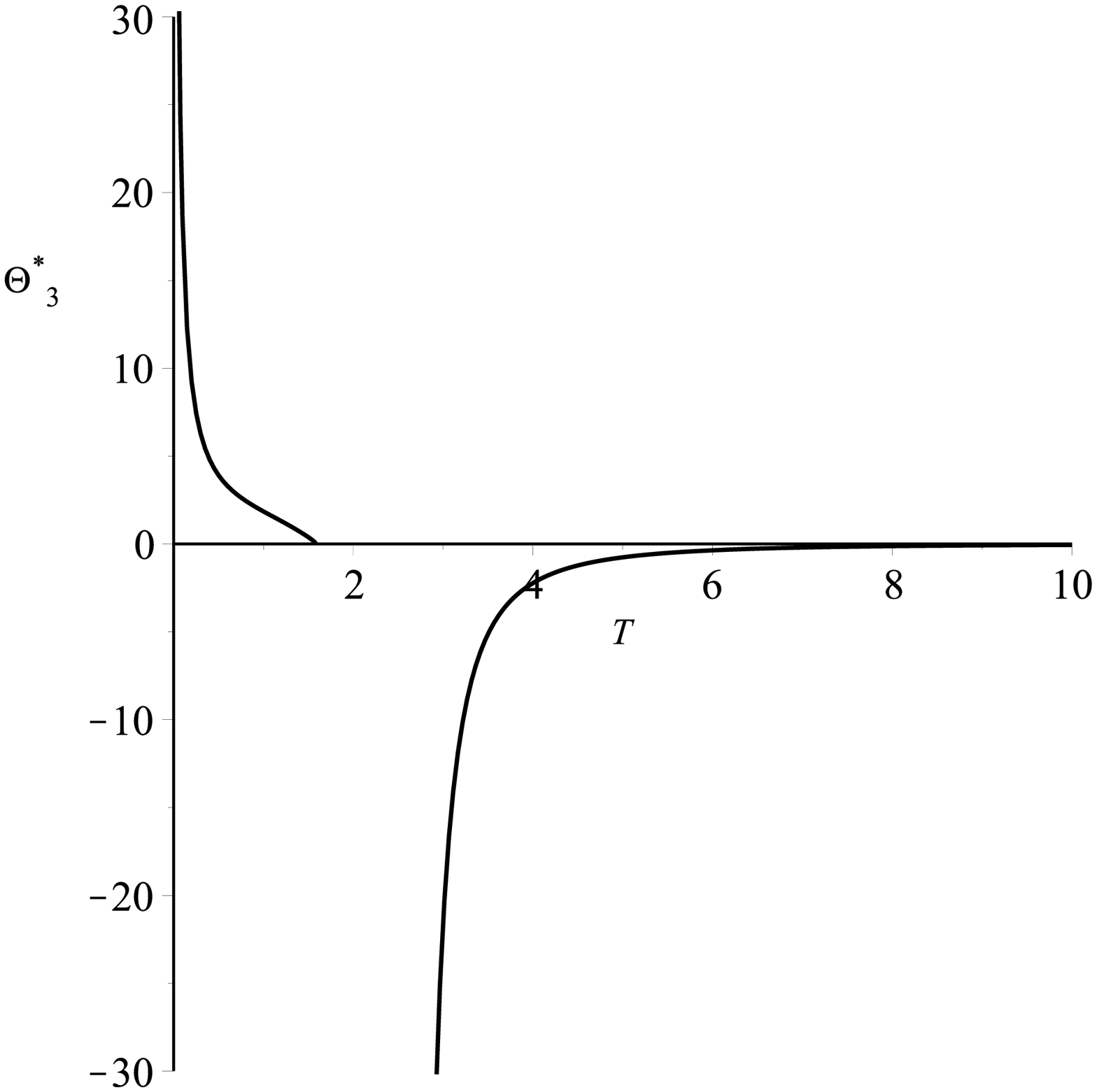}
\caption{\label{fig:epsart}  Diagram of the equation (87) is
plotted against $T.$ }
\end{figure}
\begin{figure}
\includegraphics[width=7cm]{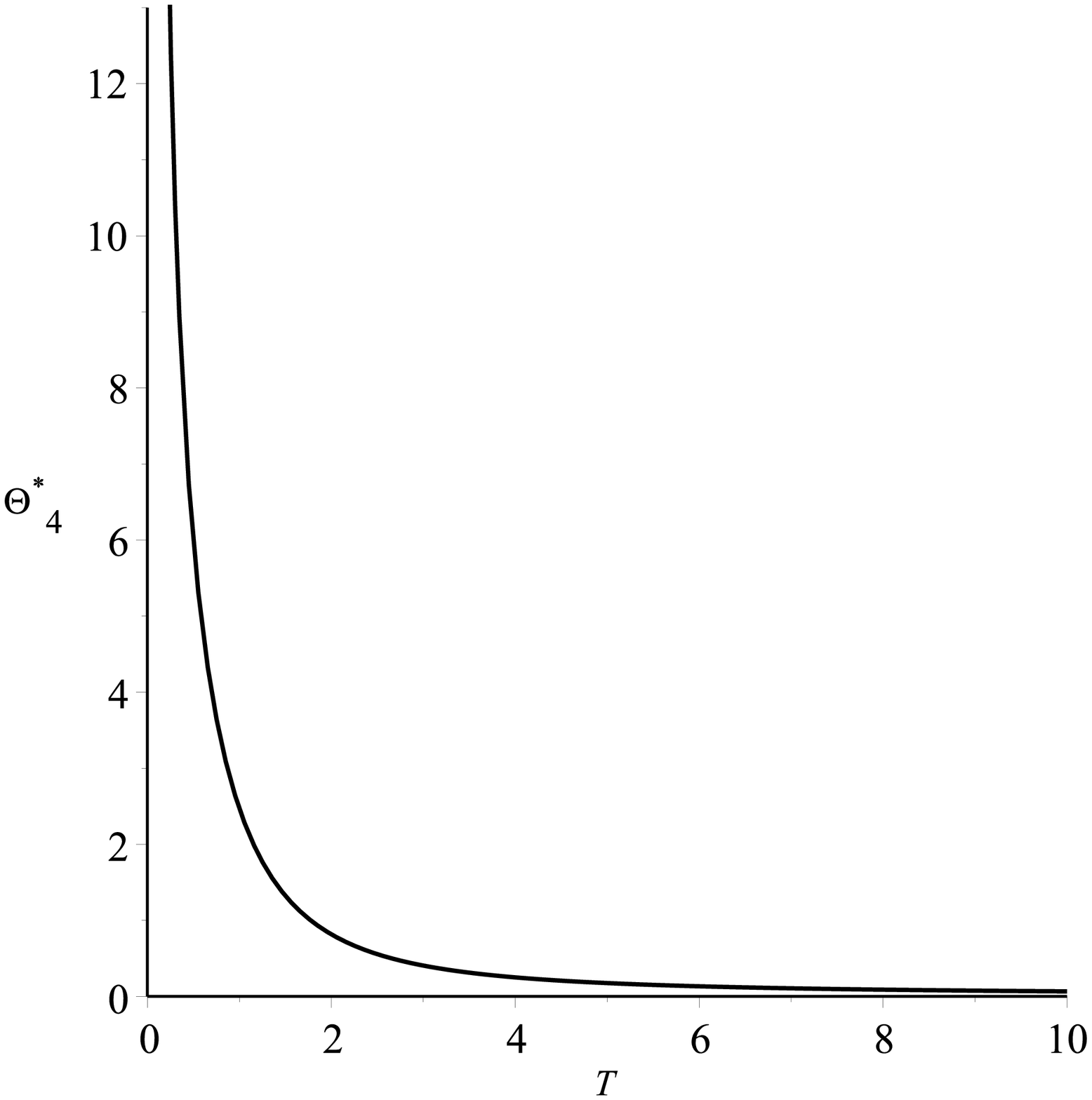}
\caption{\label{fig:epsart}  Diagram of the equation (88) is
plotted against $T.$  }
\end{figure}
\begin{figure}
\includegraphics[width=7cm]{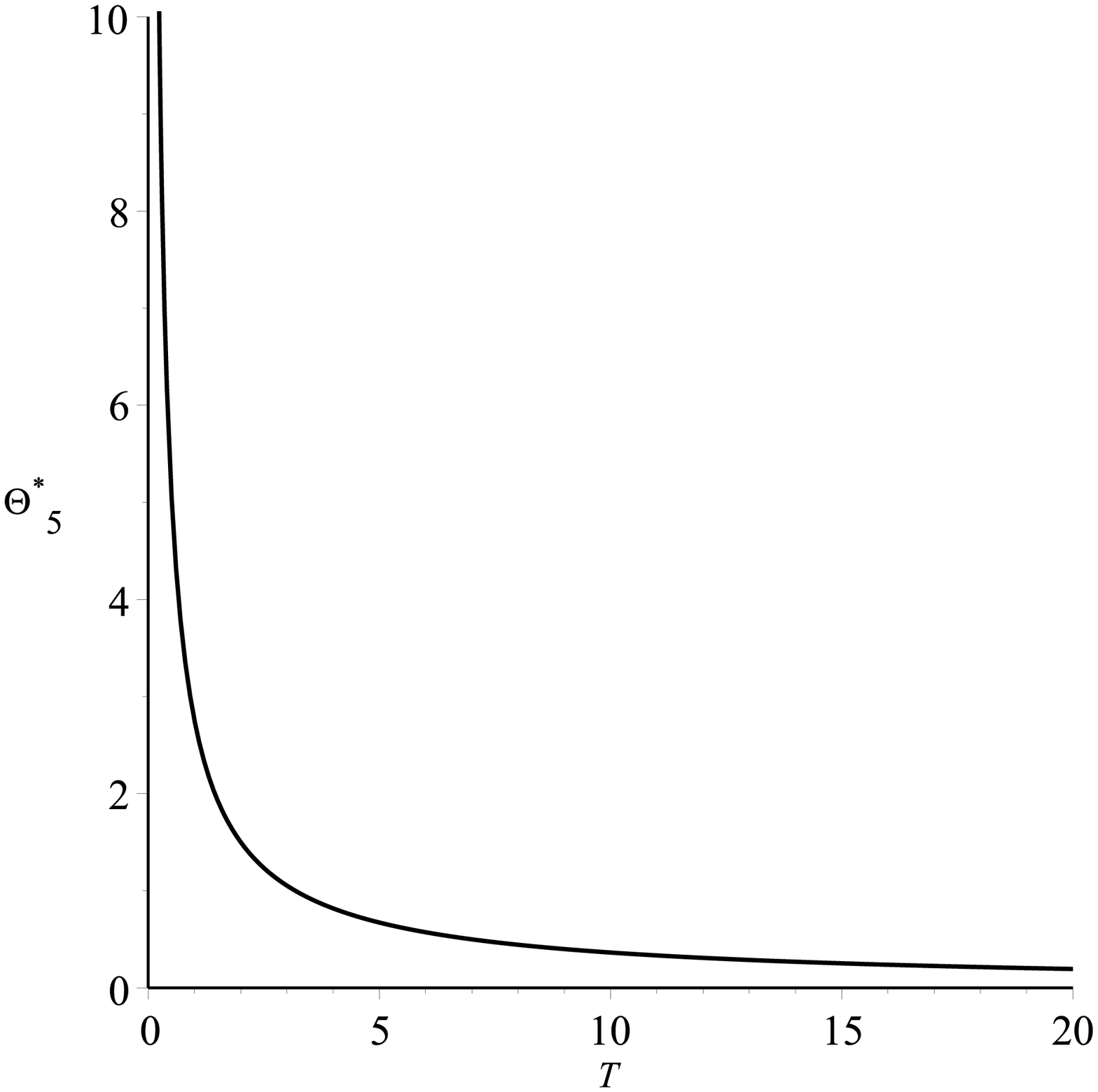}
\caption{\label{fig:epsart}  Diagram of the equation (89) is
plotted against $T.$ }
\end{figure}
\begin{figure}
\includegraphics[width=7cm]{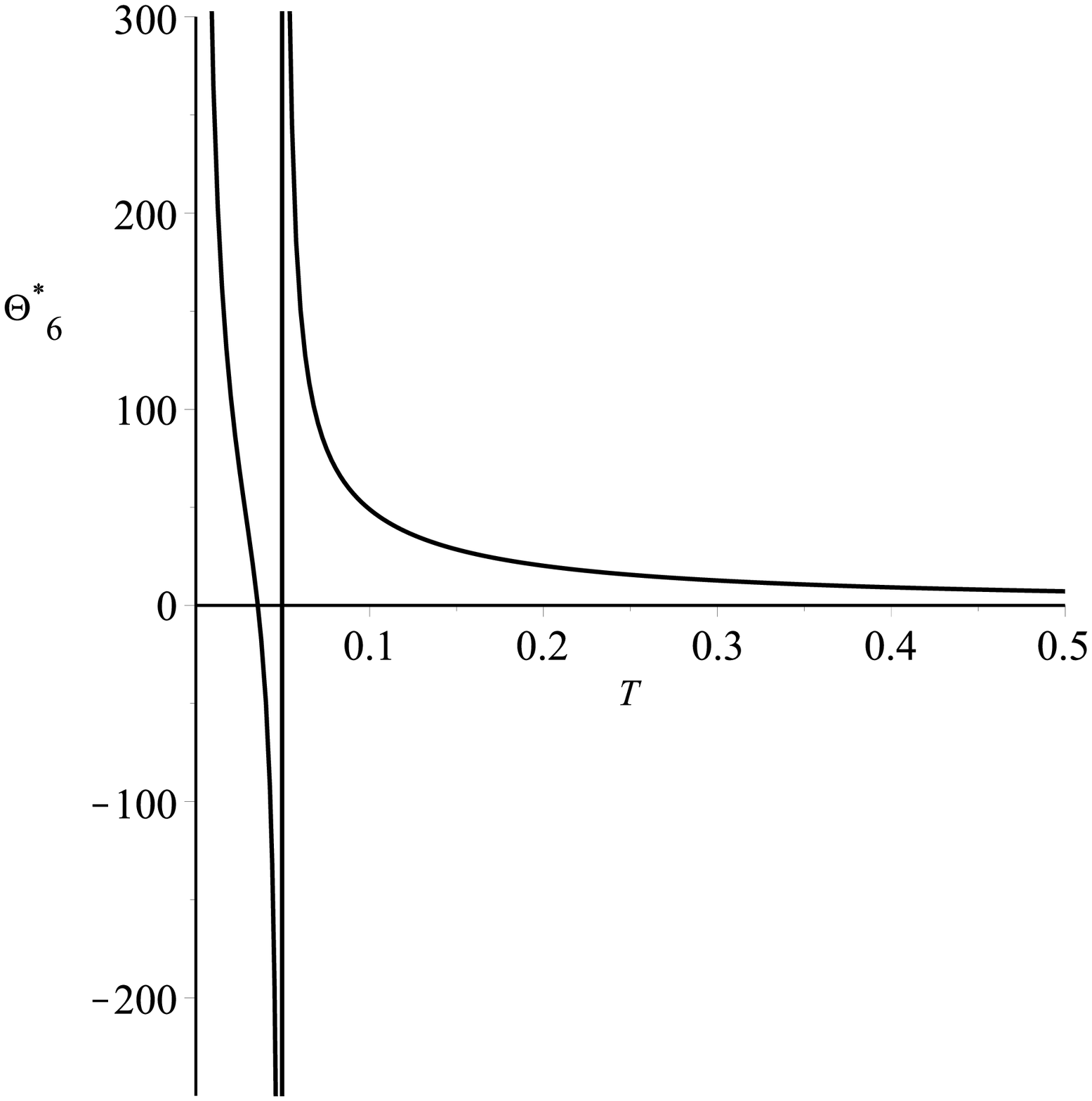}
\caption{\label{fig:epsart}  Diagram of the equation (90) is
plotted against $T.$ }
\end{figure}
\begin{figure}
\includegraphics[width=7cm]{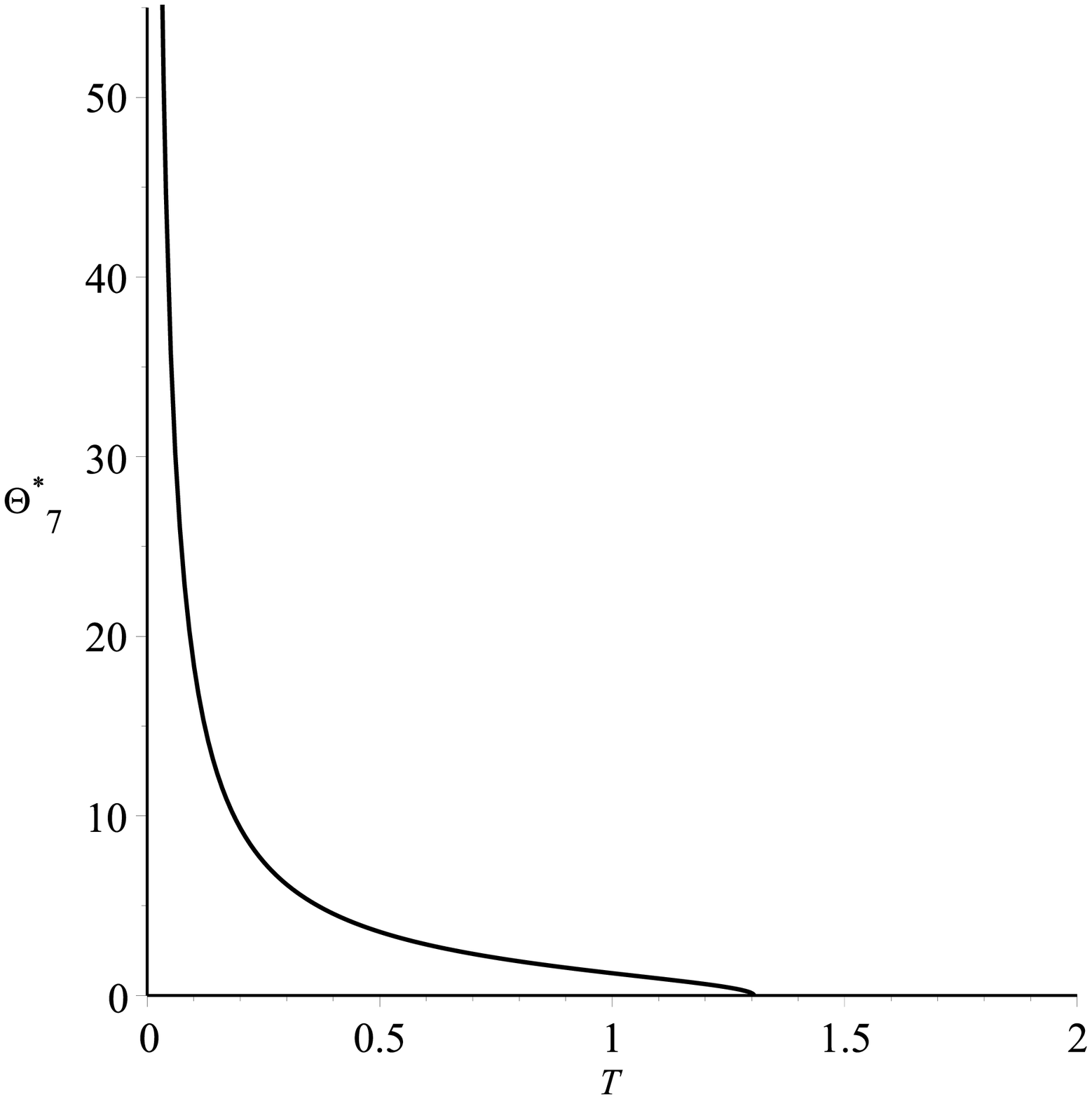}
\caption{\label{fig:epsart}  Diagram of the equation (91) is
plotted against $T.$  }
\end{figure}
\begin{figure}
\includegraphics[width=7cm]{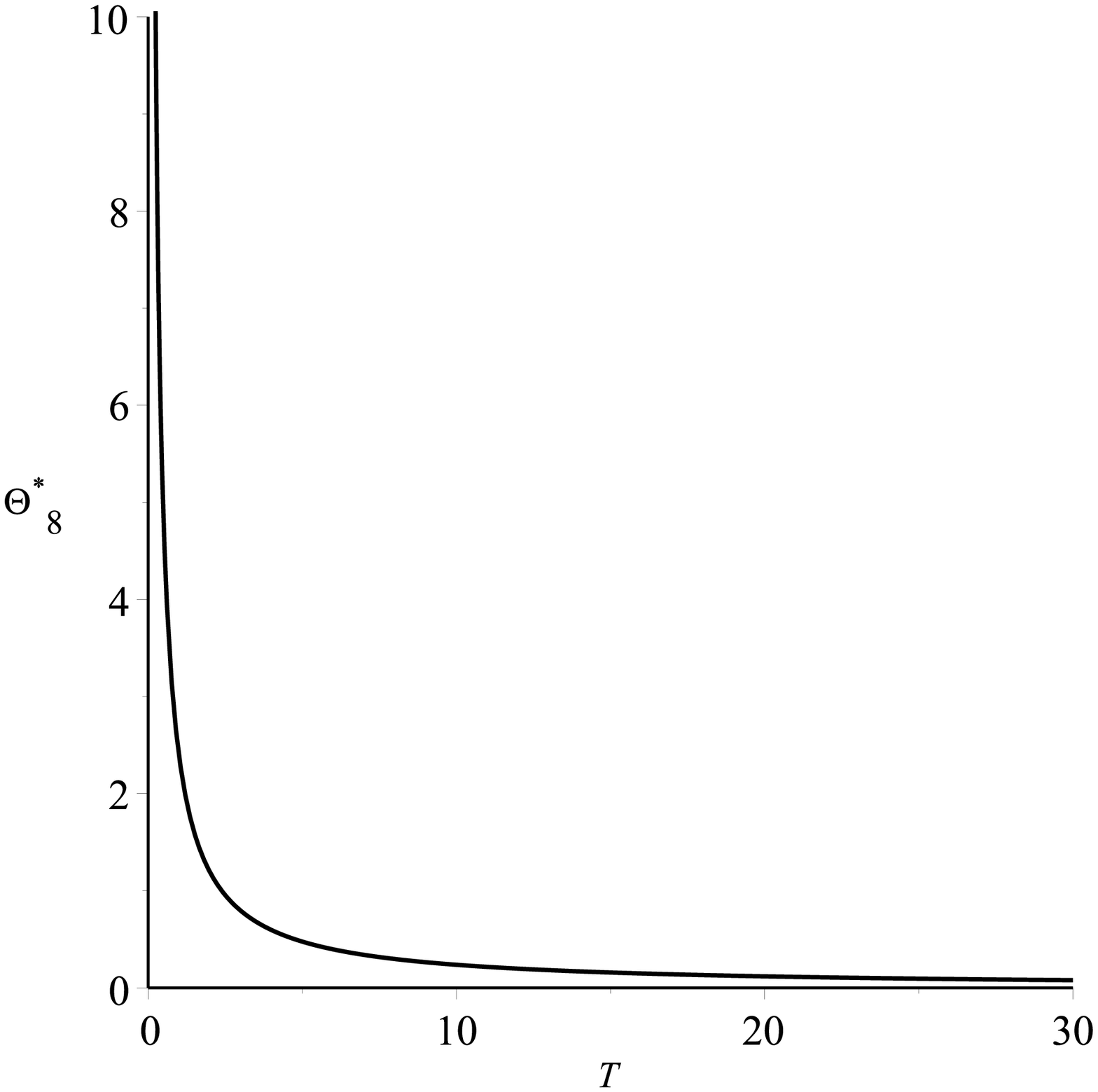}
\caption{\label{fig:epsart}  Diagram of the equation (92) is
plotted against $T.$}
\end{figure}
\begin{figure}
\includegraphics[width=7cm]{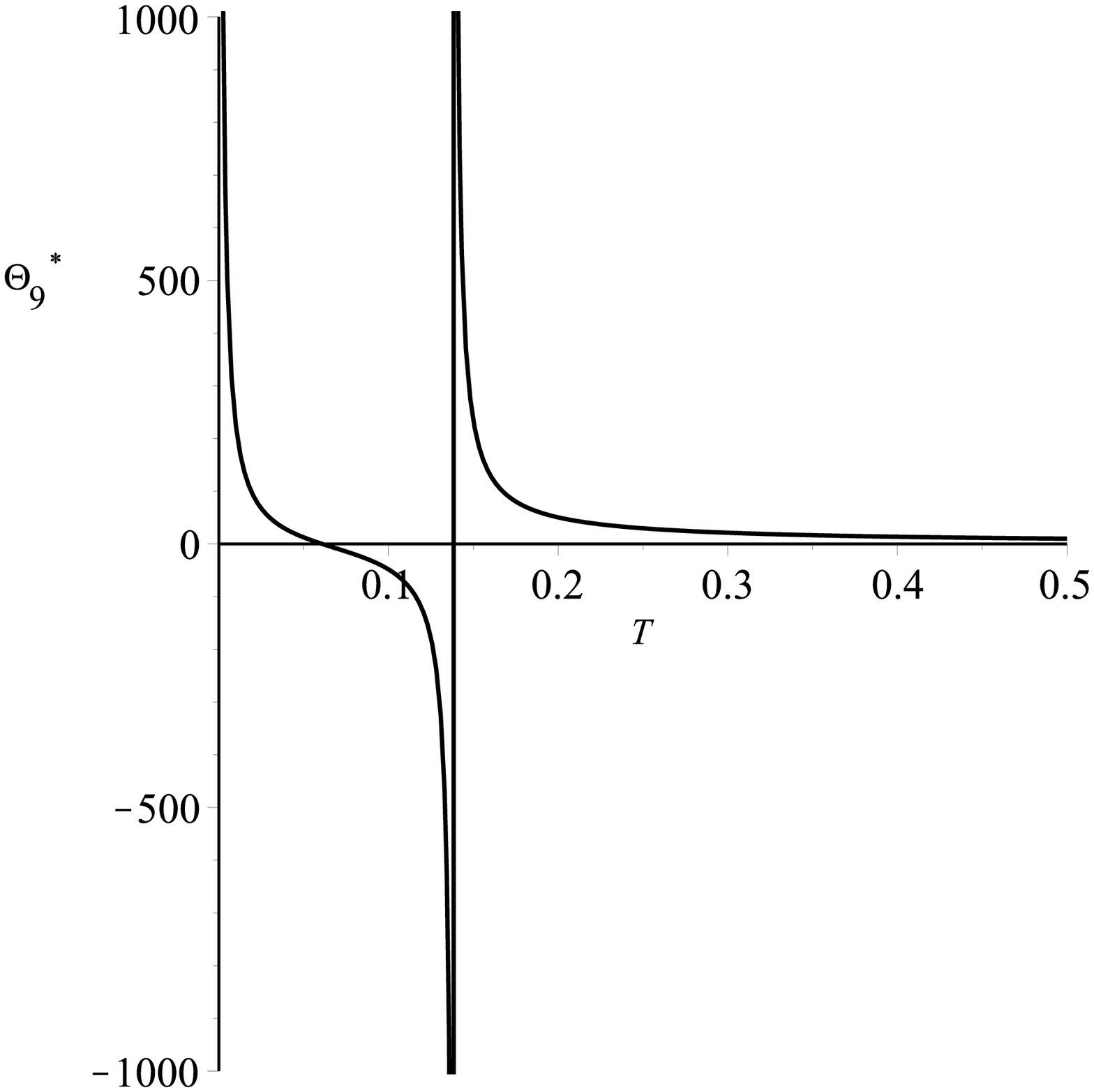}
\caption{\label{fig:epsart}  Diagram of the equation (93) is
plotted against $T.$  }
\end{figure}
\begin{figure}
\includegraphics[width=7cm]{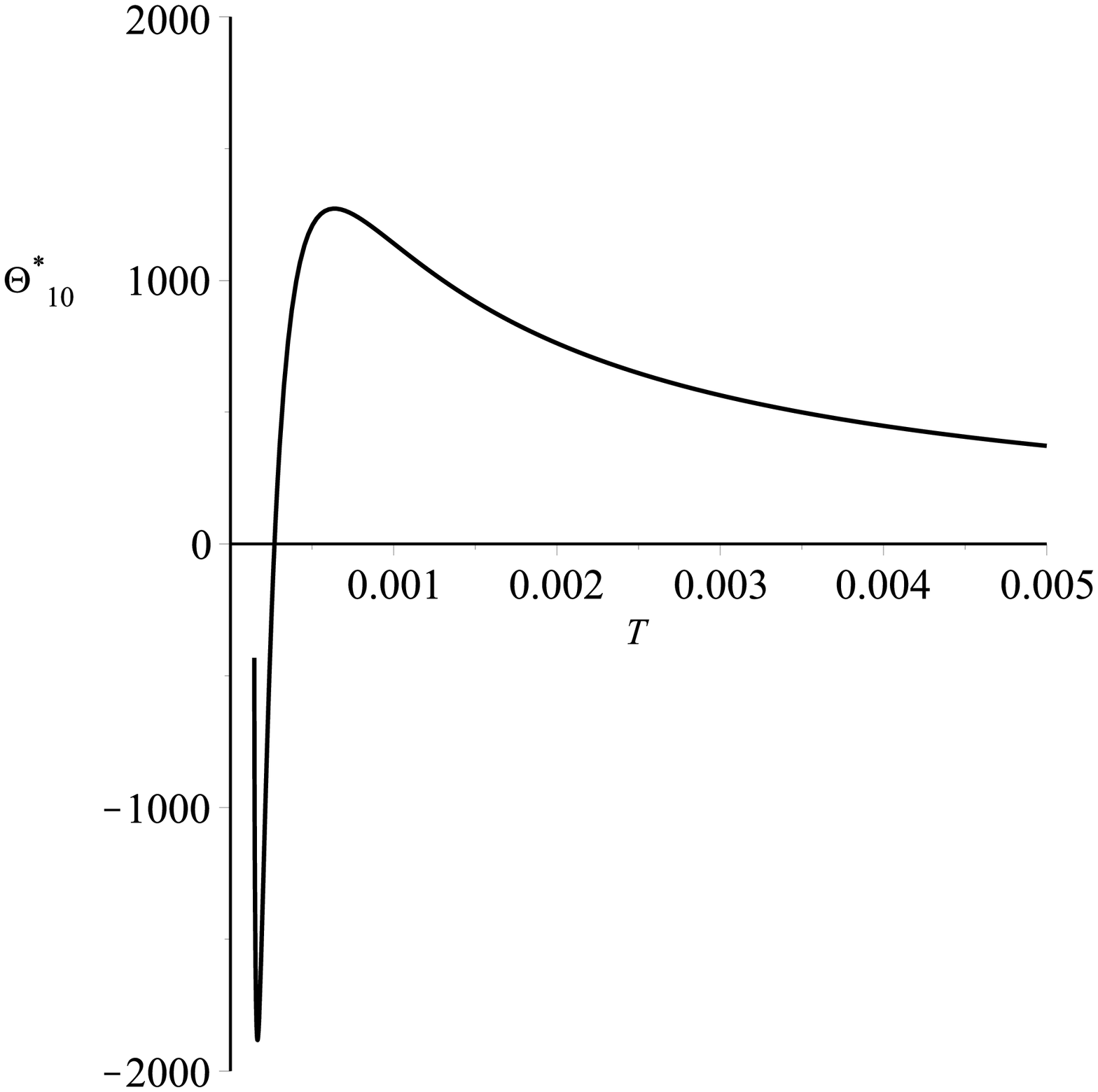}
\caption{\label{fig:epsart}  Diagram of the equation (94) is
plotted against $T.$ }
\end{figure}
\begin{figure}
\includegraphics[width=7cm]{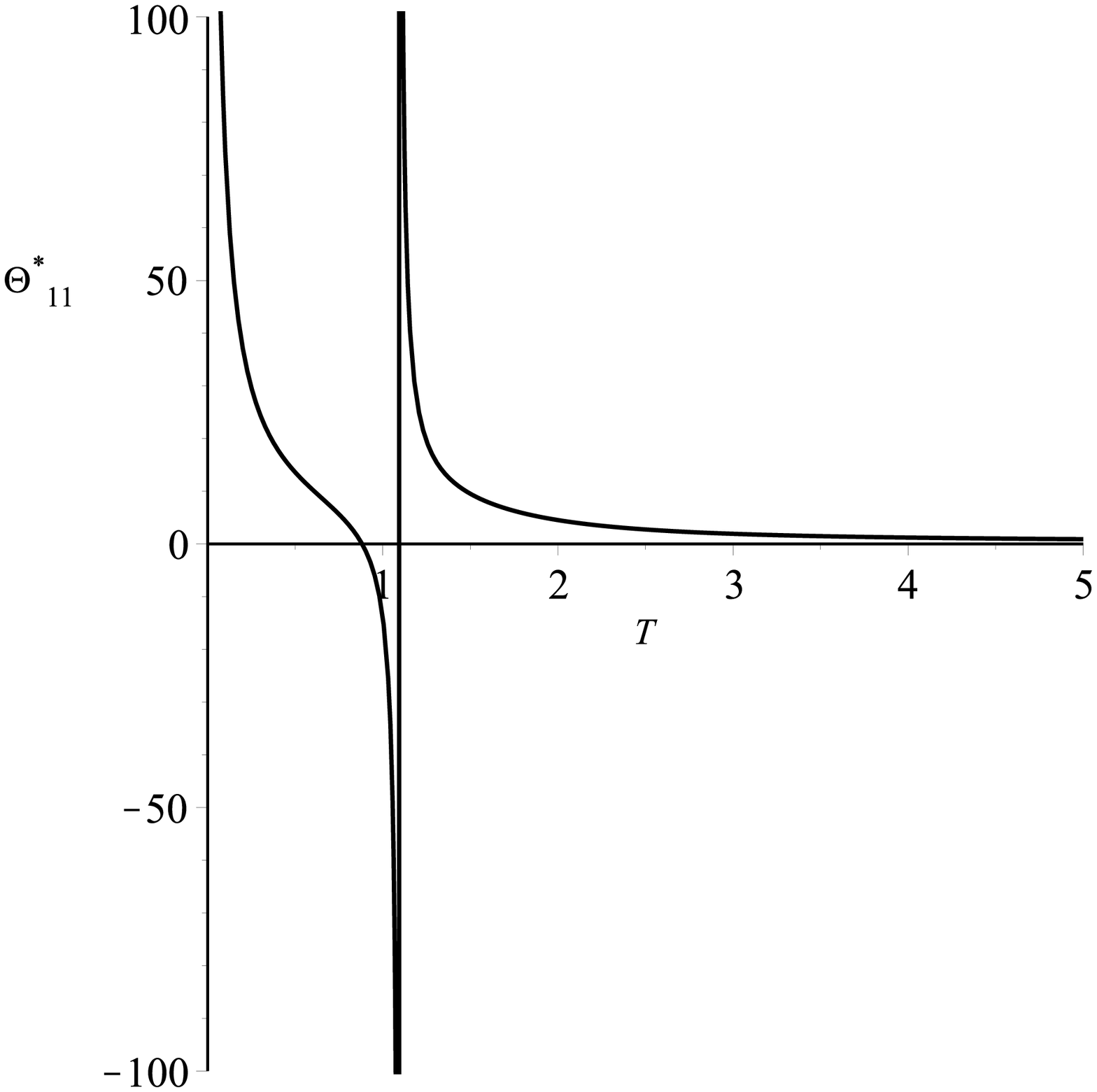}
\caption{\label{fig:epsart}  Diagram of the equation (95) is
plotted against $T.$ }
\end{figure}
\begin{figure}
\includegraphics[width=7cm]{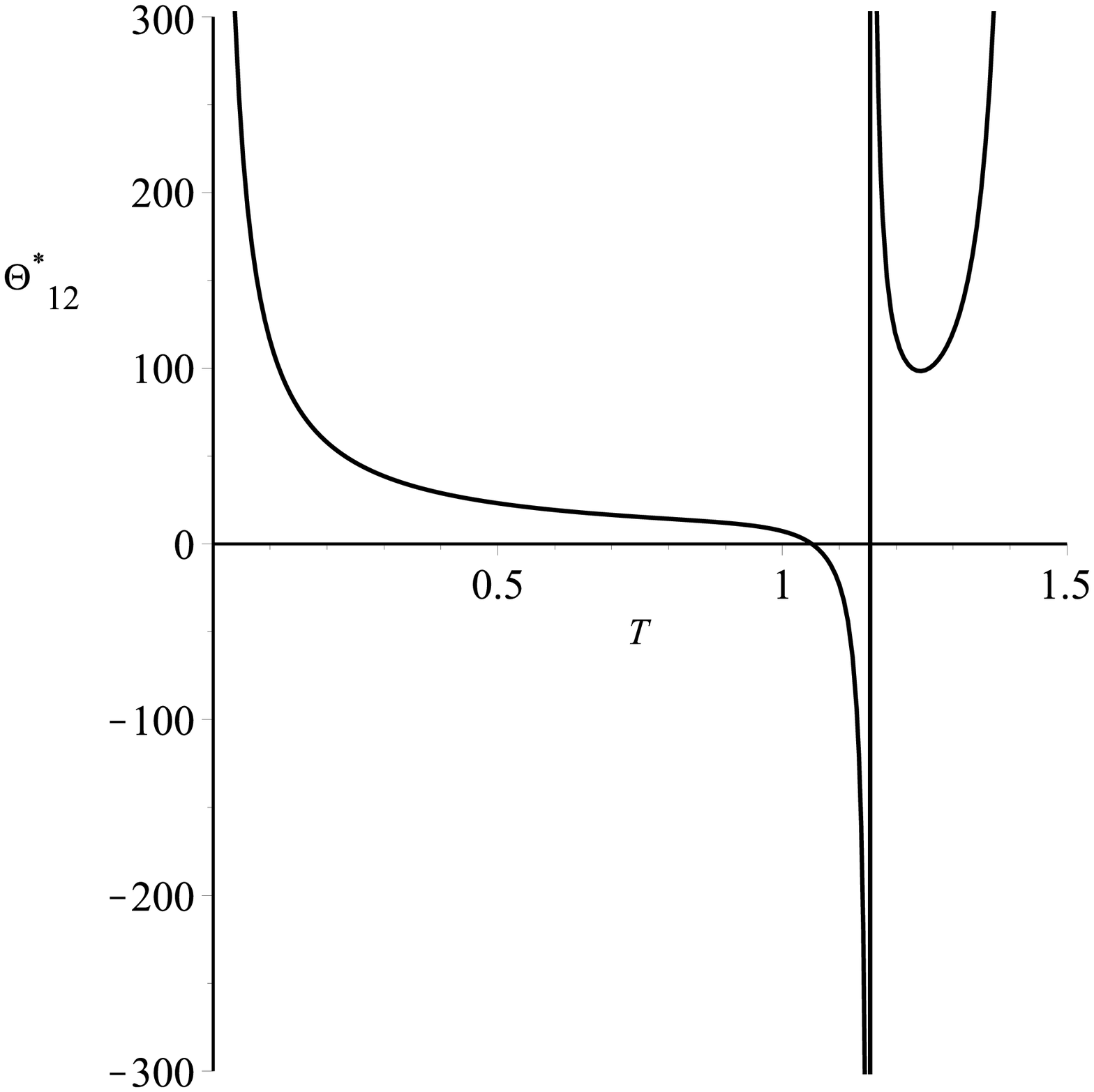}
\caption{\label{fig:epsart}  Diagram of the equation (96) is
plotted against $T.$ }
\end{figure}
\bibliography{apssamp}
\vskip 0.5cm \newpage\center \textbf{REFERENCES}
\begin{description}
\item[1] D. Lovelock. Aequationes Mathematicae, 4(1-2),127, 138, (1970),\\
 doi:10.1007/BF01817753.
\item[2] D. Lovelock. J. of Math. Phys., 12(3),498,501 (1971), doi:10.1063/1.1665613.
\item [3] N. D. Birrell and P. C. W. Davies,\textit{Quantum Fields in Curved
space}(Cambridge, England, 1982).
\item [4] N. Straumann, \textit{General
Relativity}, (Springer-Verlag Berlin Heidelberg 2004).
 \item[5] B. Fauser, J.Tolksdorf and E. Zeidler \textit{Quantum Gravity} ``Mathematical Models and
Experimental Bounds``,( Birkh\"{a}user Verlag, P.O.Box 133,
CH-4010 Basel, Switzerland, 2007).
\item [6] A. A. Starobinsky, Phys. Let. B16, 953 (1980).
\item [7] V. M\"{u}ller, H. J. Schmidt and A. A. Starobinsky,
Phys. Let. B202, 198 (1988).
\item [8] S. W. Hawking and J. C. Luttrell, Nucl. Phys. B247, 250
(1984).
\item [9] U. Kasper, Class. Quantum Grav. 10, 869 (1993).
\item [10] L. O. Pimentel and O. Obreg\'{o}n, Class. Quantum
Grav. 11, 2219 (1994).
\item [11] H. Elst van, J. E. Lidsey and R. Tavakol, Class. Quantum
Grav. 11, 2483 (1994).
\item[12] W. H. Zurek and D. N. Page, Phys. Rev. D29, N4, 628 (1984).
\item [13] P. Coles and F. Lucchin, \textit{COSMOLOGY, The origin and evolution of cosmic
Structure}, (John Wiley $\&$ Sons 1997).
\item [14] E. M.
Lifshits and L. D. Landau, \textit{The classical theory of
fields}, (Pergamon press Ltd, fourth edition 1975).
\item [15] S. Weinberg, \textit{Gravitation and Cosmology}, (John
Wiley $\&$ Sons, Inc, 1972).
\item [16] P. Musgrave
and K. Lake, Class. Quant. Grav. 13 (1996) 1885-1900,
gr-qc/9510052v3.
\item [17] C. Grenon and K. Lake, Phys. Rev. D 84, 083506 (2011),  gr-qc/1108.6320.
\item [18] R. Doran, F. S. N. Lobo and P. Crawford, Found. Phys. 38, 160
(2008), gr-qc/0609042.
\item[19] S. W. Hawking and G. F. R. Ellis, \textit{The large scale structure of space time
}, (Cambridge University Press, Cambridge, England, 1973).
\item [20] T. Harada, H. Iguchi and L. I. Nakao, gr-qc/0204008
(2002)
\item [21] C. W. Misner and D. H. Sharp, Phys. Rev. 136, 571
(1964). \end{description}
\end{document}